\documentclass[transmag]{IEEEtran}
\usepackage{latexsym}
\usepackage{graphicx}
\usepackage{amsfonts,amssymb,amsmath}
\usepackage{hyperref}
\def\BibTeX{{\rm B\kern-.05em{\sc i\kern-.025em b}\kern-.08em T\kern-.1667em\lower.7ex\hbox{E}\kern-.125emX}}

\usepackage{breqn}
\usepackage{algorithm}
\usepackage{algorithmic}
\usepackage{array}
\usepackage{url}
\usepackage{placeins}

\usepackage{multicol, blindtext}
\usepackage{multirow}
\usepackage{balance}
\usepackage{cite}
\usepackage{amsthm}
\usepackage{subcaption}
\usepackage{caption}
\usepackage{graphicx}

\begin{document}

\title{Design and Convergence Analysis of an IIC-based BICM-ID Receiver for FBMC-QAM Systems}

\author{Sumaila~Mahama, \IEEEmembership{Student Member, IEEE}, Yahya~J.~Harbi, Alister~G.~Burr,~\IEEEmembership{Senior Member,~IEEE} \\ and David~Grace,~\IEEEmembership{Senior Member,~IEEE}
\thanks{This work was supported in part by H2020-MSCA-SPOTLIGHT under Grant 722788.}
\thanks{Part of this work was presented at the 2019 European Conference on Networks and Communications (EuCNC) in Valencia, Spain \cite{Mahama:19}.}
\thanks{ Sumaila~Mahama, Alister~G.~Burr and David~Grace are with the Department of Electronic Engineering, University of York, Heslington, York, U.K. email: \{sumaila.mahama, alister.burr, david.grace\}@york.ac.uk.}
\thanks{Yahya. J. Harbi is with the University of Kufa, Najaf, Iraq email: \{yahyaj.harbi@uokufa.edu.iq\}}
\thanks{Corresponding author: Sumaila Mahama (sumaila.mahama@york.ac.uk)}
}

\IEEEtitleabstractindextext{\begin{abstract}
Next generation cellular networks are expected to connect billions of devices through its massive machine-type communication (mMTC) use case. To achieve this, waveforms with improved spectral confinement and high spectral efficiency compared to cyclic-prefix orthogonal division multiplexing (CP-OFDM) are required. In this paper, filter-bank muilticarrier with quadrature amplitude modulation (FBMC-QAM) is studied as an alternative to CP-OFDM for such applications. However, FBMC-QAM presents the challenge of high intrinsic interference due to the loss of complex orthogonality in time-frequency domain. Bit-interleaved coded modulation with iterative decoding (BICM-ID) has been shown to exhibit capacity-approaching decoding performance. Therefore, in this paper, an iterative interference cancellation (IIC) based BICM-ID receiver is designed to cancel the intrinsic interference in FBMC-QAM systems. The proposed receiver consist of an inner decoder, which combines iterative demodulation and interference cancellation, and an outer decoder which is a low density parity check (LDPC) channel decoder. To evaluate the convergence behaviour of the proposed receiver, extrinsic information transfer (EXIT) chart analysis is employed, where EXIT curves for the components of the iterative decoder are derived. Numerical results show that the intrinsic interference in FBMC-QAM systems can be removed by adopting the proposed IIC-based BICM-ID receiver. With a receiver that is able to successfully cancel intrinsic interference, FBMC-QAM becomes an interesting alternative waveform for asynchronous mMTC applications due to its superior frequency localization compared to CP-OFDM. Furthermore, it has been shown that FBMC-QAM with the IIC-based receiver can achieve similar bit-error-rate (BER) performance as a CP-OFDM benchmark under different fading channels. Finally, the complexity of the proposed receiver for FBMC-QAM is analysed and compared to the complexity of the CP-OFDM benchmark. 
\end{abstract}


}

\maketitle

\section{INTRODUCTION}
\IEEEPARstart{C}{ellular} communication networks have evolved over the past few decades. From a network that was originally designed for voice calls, current forth generation (4G) networks can support broadband applications with huge data demands. The fifth generation of mobile networks, 5G, is expected to support additional network functionalities, such as low latency communication and massive Machine-Type Communication (mMTC). For mMTC scenarios, for example Internet of Things (IoT), billions of devices will connect and interact with each other and with humans through the internet \cite{Akpakwu:2018}. To connect all these devices to the network, the physical layer waveform must be able support different service types, be spectrally efficient and enable asynchronous transmissions \cite{Chen:18}. 

Although orthogonal frequency division multiplexing using cyclic prefix (CP-OFDM) has been adopted as the standard multicarrier modulation scheme for many wireless technologies, some of its drawbacks make CP-OFDM not the best choice of waveform for future mMTC applications. For example, the use of rectangular pulse shaping filters in CP-OFDM results in high out-of-band (OOB) emissions. This causes severe interference to adjacent frequency bands which may be allocated to other mMTC nodes. Therefore, to avoid interference, CP-OFDM requires strict synchronization in order to maintain orthogonality among subcarriers. With mMTC aiming to connect billions of devices, the synchronization overhead (CP and guard band) associated with CP-OFDM may be unmanageable. Therefore, to support the various emerging applications of future communication networks, new waveforms that are spectrally efficient and more suitable for asynchronous access have been investigated recently \cite{MPischella:18,CSexton:18,TXu:18}.  
 
One modulation scheme that has gained significant research attention for asynchronous mMTC applications is filter-bank multicarrier (FBMC) modulation \cite{Bellanger:10,Sahin:14}. FBMC utilizes an advanced prototype filter on each subcarrier, resulting in improved time-frequency localization. This results in a considerable decrease in the OOB leakage to adjoining frequency bands. Also, in contrast to CP-OFDM which depends on orthogonality to recover transmitted signals, FBMC relies on its good frequency localization for transmit signal reconstruction \cite{Chen:18}. Therefore, strict synchronization among subchannels is not needed. This makes FBMC systems suitable for asynchronous transmissions and significantly reduces the overhead due to synchronous communication in CP-OFDM. Moreover, FBMC systems are robust against channel frequency selectivity without using any CP, which improves the spectral efficiency compared to CP-OFDM.  

As stated in the Balian-Low Theorem, multicarrier systems cannot achieve orthogonality, high time-frequency localization and maximum symbol density simultaneously \cite{Demmer:19,Benedetto:94}. Consequently, the improved spectral confinement of FBMC is obtained at the expense of the loss of orthogonality between subcarriers. To deal with the non-orthogonality problem of FBMC, offset quadrature amplitude modulation (OQAM) was proposed in \cite{Bellanger2010}. However, FBMC-OQAM only achieves orthogonality in the real domain and suffers from intrinsic interference in complex fading channels. This limits the direct implementation of conventional LTE-based schemes such as channel estimation, space-time block code (STBC) and maximum likelihood detection (MLD) \cite{Zakaria:12}. To resolve the problems in FBMC-OQAM, the transmission of complex QAM symbols in FBMC systems was proposed in \cite{Zakaria:14,HNam:16}.

By transmitting QAM signals, the FBMC systems cannot guarantee complex domain orthogonality, resulting in intrinsic interference. To restore orthogonality in FBMC-QAM, the authors in \cite{HNam:16} proposed a twin-filter model that together satisfy complex orthogonality in ideal channels. However, in fading channels, residual interference remains. Also, the filter design cause filter coefficient discontinuities which result in higher OOB emissions compared to CP-OFDM. FMBC-QAM systems with optimized filter sets have been proposed to find a trade-off between satisfying orthogonality and maintaining a good time-frequency confinement \cite{Yun:15,DonSim:19}. In \cite{DonSim:19}, performance gains in bit-error-rate (BER) and OOB emission is achieved at the cost of a highly complex equalizer compared to CP-OFDM. A linearly processed FBMC-QAM system was proposed in \cite{Jintae:18} which achieves similar OOB emission as FBMC-OQAM. However, the proposed scheme transmits data only on even-numbered subcarriers, thereby making use of only 50 percent of the available frequency resources. 
 
Alternatively, iterative detection and decoding (IDD) has been shown to approach the optimal performance of different communication systems by iteratively performing independent detection and decoding of the received signal \cite{QLi:16,QLi:14}. The IDD receiver consist of a soft-input soft-output (SISO) demapper and a channel decoder that iteratively exchange “soft” information to improve detection performance. Bit-interleaved coded modulation (BICM) based iterative decoding, BICM-ID, was introduced in \cite{XLi:97} to achieve near capacity performance at a lower complexity. Some recent works have studied the combination of BICM-ID and iterative interference cancellation (IIC) to remove intrinsic interference in FBMC systems \cite{Zakaria:14,YahyaJHarbi:16,YahyaJHarbi:18,SMahama:19,Mahama:19}. The works in \cite{Zakaria:14} and \cite{SMahama:19} studied the use of IIC receivers for uncoded FBMC-QAM systems. In both papers it was shown that IIC can reduce the intrinsic interference power in FBMC-QAM systems. However, in frequency selective fading channels an error floor appears in the BER performance at high signal-to-noise ratio (SNR). \cite{Caus:16} investigated the application of non-binary LDPC codes to FBMC-OQAM and showed that existing adaptive coding and modulation schemes can be easily modified to accommodate the properties of non-binary LDPC and FBMC-OQAM. However, the work in \cite{Caus:16} does not tackle the intrinsic interference problem in FBMC-OQAM. An iterative detection and decoding (IDD) receiver based on LDPC decoding was proposed by two of the present authors in \cite{YahyaJHarbi:16} and \cite{YahyaJHarbi:18} for single antenna FBMC-OQAM and multiple-input-multiple-output (MIMO) FBMC-OQAM, respectively. Nonetheless, as mentioned above, the implementation of FBMC-OQAM in some MIMO applications, such as STBC, is challenging and hence in \cite{Mahama:19} we extended these iterative approaches to FBMC-QAM. \cite{Mahama:19} studied the performance of an IIC-based IDD receiver for FBMC-QAM based Narrowband IoT (NB-IoT) systems. The results showed that the IDD receiver can significantly improve the BER performance of FBMC-QAM systems by iteratively removing the intrinsic interference. However, as with all iterative receivers, the complexity and convergence behaviour are two key design requirements.

In this paper, we study an IIC-based BICM-ID receiver for FBMC-QAM systems with a single prototype filter, under different fading channels. To overcome the intrinsic interference introduced by the loss of orthogonality the proposed receiver combines two component decoders. The received signal is fed to the \textit{inner decoder}, which performs FBMC-QAM demodulation and demapping and passes the soft information of coded bits at the ouput of the demapper to the \textit{outer decoder}. The outer decoder (LDPC decoder) performs iterative decoding by passing the received information between its variable node decoder (VND) and check node decoder (CND) for a predefined number of iterations. The signal output by the outer decoder is again fed back to the inner decoder for interference estimation and cancellation. Thus the intrinsic interference is mitigated following a scheduled number of iterations of the proposed receiver.  

In order to investigate the required number of iterations to achieve a given BER performance of the proposed receiver with least complexity, we adopt the concept of extrinsic information transfer (EXIT) chart analysis \cite{StenBrink:04, MEHajjar:14, YahyaJHarbi:17, IWKang:18}. Based on the EXIT chart analysis the characteristics of the iterative receiver can be predicted for different performance requirements. For example, EXIT chart analysis is implemented in \cite{IWKang:18} to predict the optimum row-permutation interleavers of twin-interleaver BICM-ID systems. Again, an EXIT chart approach based on trellis search was introduced in \cite{QLi:16} to determine the loop schedule that achieves a target BER performance with minimized complexity. Note that in the preliminary results presented in \cite{Mahama:19}, an IDD receiver is proposed for FBMC-QAM systems. However, the iterative convergence behaviour using EXIT charts was not discussed. In this paper, as shown in subsection \ref{SubSecResults1}, EXIT chart analysis will enable the prediction of the number of iterations needed in the two main loops of the proposed receiver: i.e. (i) the number of iterations between the VND and CND of the outer decoder and (ii) the number of iterations between the inner decoder and the outer decoder. This in turn provides a basis for estimating the complexity of the proposed receiver. The main contributions of this paper are summarized as follows:
\begin{itemize}
\item The use of an IIC-based BICM-ID receiver for FBMC-QAM systems is analysed. First, an analytical expression is derived for the received FBMC-QAM signal in terms of the desired signal, inter-carrier interference (ICI), inter-symbol interference (ISI) and noise. The ICI and ISI represent the intrinsic interference caused by the loss of complex orthogonality in FBMC-QAM.  
\item Based on the derived interference model, we estimate and cancel the intrinsic interference by subtracting the estimated interference during each iteration of the proposed receiver. 
\item To evaluate the convergence behaviour of the proposed receiver, the EXIT functions of the components of the iterative receiver are derived. In numerical simulation results, the number of iterations and the convergence threshold of the proposed receiver under different fading channels is evaluated. Also, we show that the proposed receiver can significantly improve the BER performance of FBMC-QAM systems. 
\item Finally, we analyse and compare the computational complexity of the proposed receiver for FBMC-QAM and CP-OFDM.     
\end{itemize}

The remainder of this paper is organized as follows. Section II describes the FBMC-QAM system model and the structure of the proposed IIC-based BICM-ID receiver. The convergence behaviour of the proposed iterative receiver using EXIT-chart analysis is presented in Section III. Performance and complexity evaluation of the BICM-ID receiver for different frequency-selective channel models are illustrated in Section IV. Finally, conclusions are drawn in Section V.

\begin{figure*}
\centering
\includegraphics[width=6in]{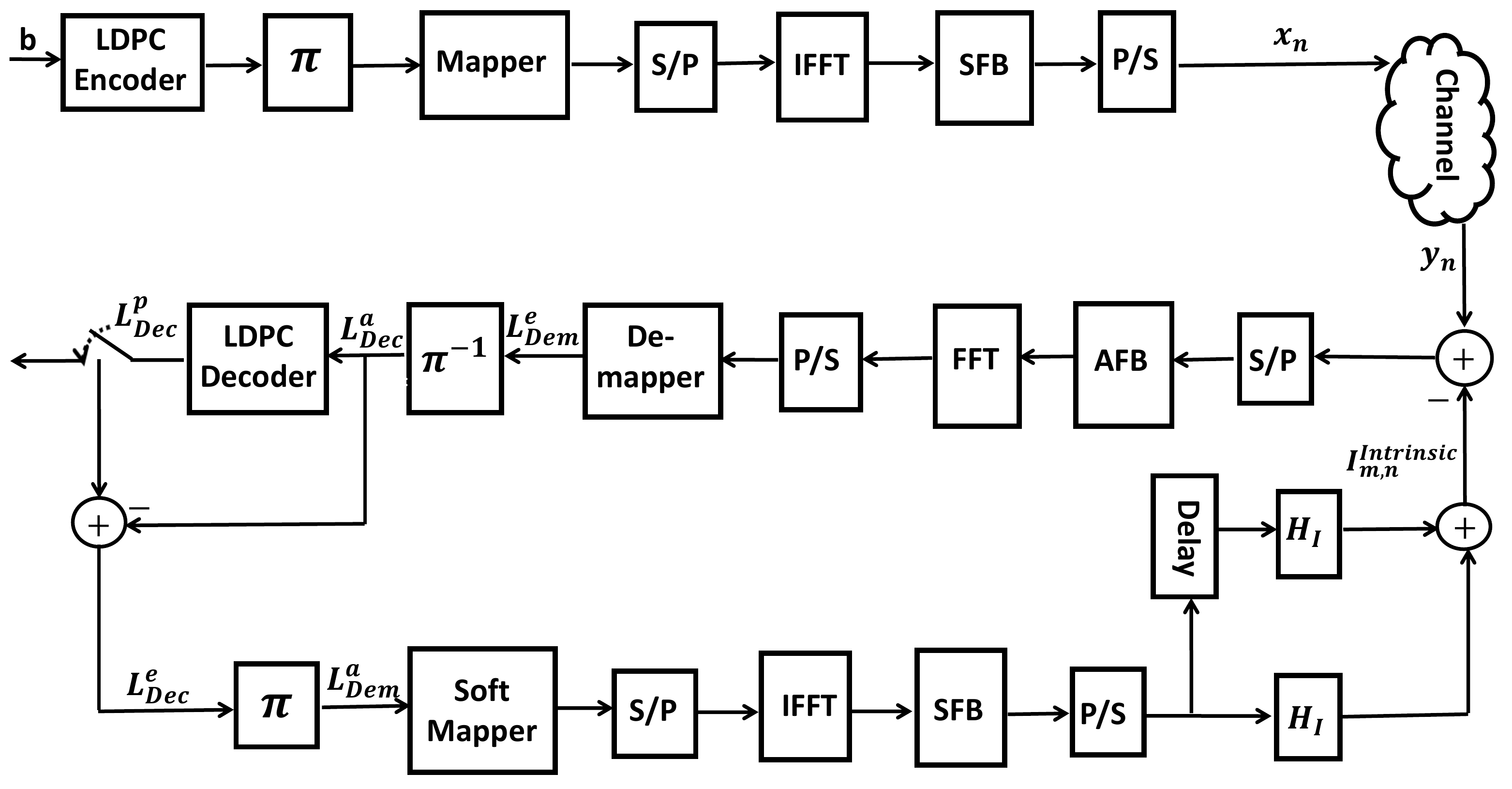}
\caption{Block diagram of transmitter and BICM-ID receiver}
\end{figure*}

\begin{figure*}
\centering
\includegraphics[width=6in]{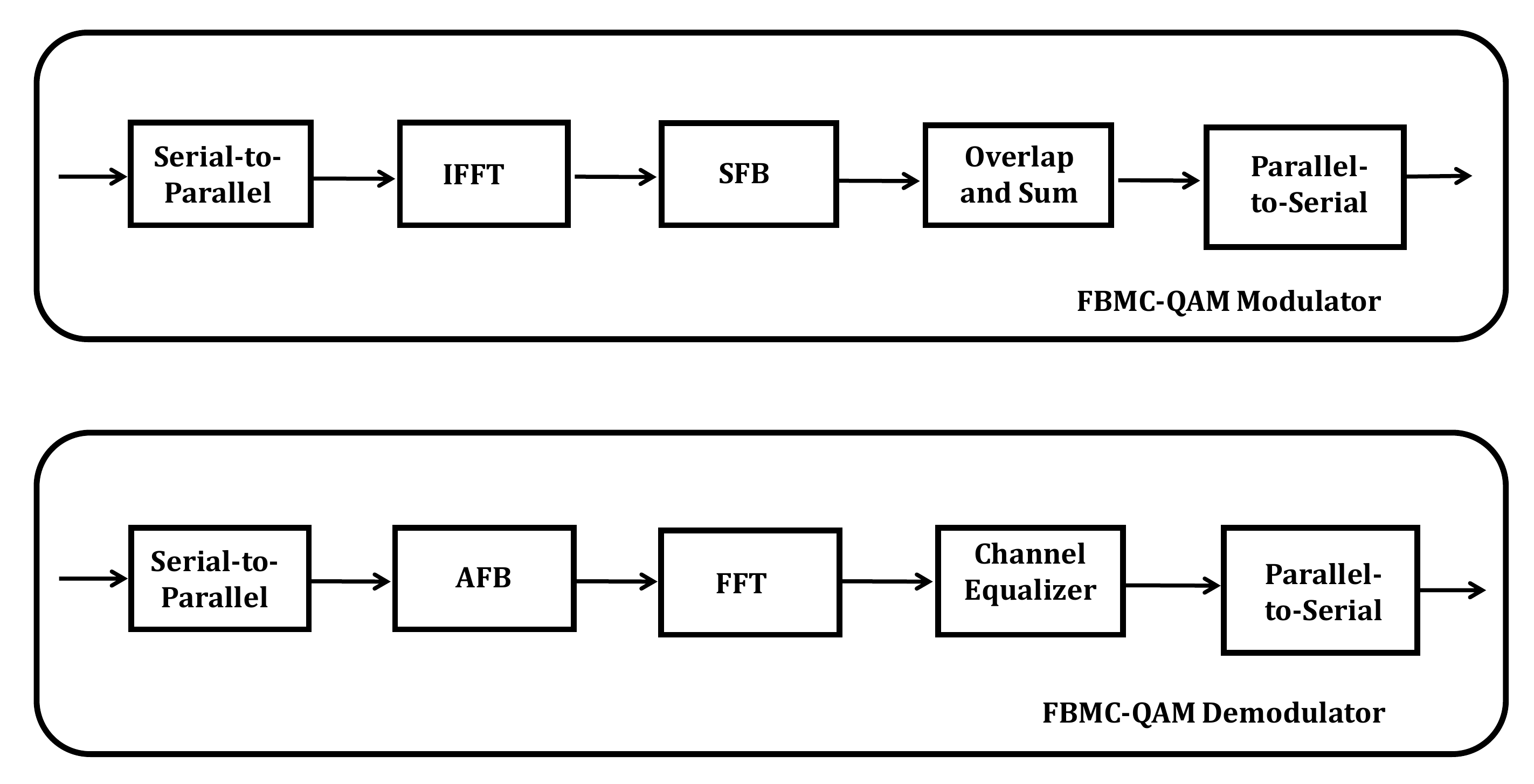}
\caption{FBMC-QAM modulator and demodulator}
\end{figure*}

\section{System Model} 
\label{SysModel} 
\subsection{FBMC-QAM System Model}
The transceiver structure of the proposed IIC-based BICM-ID system is shown in Fig. 1 on the next page. We consider a single antenna FBMC-QAM system. Contrary to the work in \cite{HNam:16}, a single prototype filter is used on all subcarriers to maintain the superior OOB emission property of FBMC. At the transmitter, a stream of information bits  \( \mathbf{b} = [b_1, b_2, \ldots, b_{L_b}] \) of length \(L_b\) are encoded by a channel encoder. The encoder outputs a codeword \( \mathbf{c} = [c_1, c_2, \ldots, c_{L_c}] \) of length \(L_c\) with a coding rate of \(R_c = L_b/L_c\). A low density parity check code (LDPC) encoder is considered in this paper due to its widespread application in modern communication systems, such as IEEE 802.11n and 5G new radio (NR). The encoded bits are randomly interleaved in order to randomize the burst errors and a set of \(\mathcal{M}\) interleaved bits
\(\{ a_{n}^1, a_{n}^2, \ldots, a_{n}^{\mathcal{M}} \} \) are QAM modulated to the \(n\)-th transmission symbol \(\mathbf{a}_n\), where \( a_{n}^{\mathit{q}} \in \{ \pm 1\} \) denotes the \( \mathit{q} \)-th bit of \(\mathbf{a}_n\). The resulting signal goes through the FBMC-QAM modulator, which consists of a serial-to-parallel converter, inverse fast Fourier transform (IFFT), synthesis filter bank (SFB) block and parallel-to-serial converter as shown in Fig. 2. After IFFT, the discrete time-domain FBMC-QAM signal vector associated with the \(n\)-th symbol interval is expressed as 
\begin{equation}
\label{EQ1}
\mathbf{s}_n = \mathbf{\Phi}^H \mathbf{a}_n 
\end{equation}
where  \( \mathbf{a}_{n}  = [a_{0,n}, a_{1,n}, \ldots, a_{M-1,n}]\) is the 
\( M \times 1\) symbol vector in the frequency domain before IFFT, \( a_{m,n}\) is the data at the \(m\)-th subcarrier of the \(n\)-th symbol, \(M \) is the total number of subcarriers and \(\mathbf{\Phi} \) is the \(M \times M\) unitary discrete Fourier transform (DFT) matrix whose entry on the \(l\)-th row and \(t\)-th column is \((1/\sqrt{M})  e^{-j \frac{2 \pi lt}{M}} \).  

The IFFT output is oversampled by a factor \( K \) and filtered by the synthesis filter bank (SFB). An efficient and low complexity implementation of the per-subcarrier filtering in FBMC is the poly-phase network (PPN) \cite{Jintae:18}. The signal after SFB processing is given as
\begin{equation}
  \begin{aligned}
\label{EQ2}
\mathbf{x}_n &= \textbf{G} \mathbf{s}_n \\
             &= \textbf{G} \mathbf{\Phi}^H \mathbf{a}_n 
  \end{aligned}
\end{equation}
where \( \mathbf{x}_n  = [x_{0,n}, x_{1,n}, \ldots, x_{\Gamma-1,n}]\) is the vector of the transmitted data, \(\Gamma = K \times M \) and \( \mathbf{G}\) is the PPN matrix whose \(l\)-th row and \(t\)-th column is given as \cite{Jintae:18}
\begin{equation}
\label{EQ3}
[\mathbf{G}]_{lt} = \begin{cases}
    g[(l-t)M],& \text{for } 0 \leq l-t < K \\
    0,              & \text{otherwise}
\end{cases}
\end{equation}
with \( g[i] \) as the prototype filter coefficient. The PHYDYAS prototype filter has been widely adopted for FBMC systems due to its low OOB emission performance and was shown in \cite{Zakaria:14} to be effective for interference cancellation receivers. For this reason, the PHYDYAS prototype filter is employed in this paper. Refer to \cite{Chen:18} for performance and comparison of different prototype filters used is FBMC systems. 

In the presence of a frequency-selective fading channel, the discrete-time domain FBMC-QAM signal vector at the receiver  \( \mathbf{y}_n  = [y_{0,n}, y_{1,n}, \ldots, y_{\Gamma-1,n}]\) is given by 
\begin{equation}
\label{EQ4}
\mathbf{y}_n = \mathbf{H} \mathbf{x}_n + \mathbf{z}_n
\end{equation} 
where \(\mathbf{H}\) denotes the \( \Gamma \times \Gamma \) multipath channel matrix with entries given as 
\begin{equation}
\label{EQ5}
[\mathbf{H}]_{\tau_1,\tau_2} = \begin{cases}
    h(\tau_1-\tau_2),& \text{for } 0 \leq \tau_1-\tau_2 < L_{ch}, \\
    0,              & \text{otherwise}
\end{cases}
\end{equation}
\( h(\tau) \) is the time-domain multipath channel impulse response, 
\( \tau \) is the multipath delay, \( L_{ch} \) is the length of the channel and \( \mathbf{z}_n   \) is the \( \Gamma \times 1\) vector representing the additive white complex Gaussian noise (AWGN).

The received signal is passed through the FBMC-QAM demodulator, which reverses the operations of the FBMC-QAM modulator (See Fig. 2). Ideally, the analysis filter bank (AFB) and SFB must satisfy the perfect reconstruction (PR) condition \cite{Chen:18}. However, under fading channel conditions it is impossible to achieve PR conditions. Thus, prototype filters are designed to satisfy NPR characteristics. To achieve at least NPR the SFB and AFB filters are chosen to be complex conjugate and time reverse versions of each other \cite{YahyaJHarbi:16}. Therefore, the signal at the output of AFB after receive filtering is expressed as
\begin{equation}
  \begin{aligned}
\label{EQ6}
\mathbf{y}_n^f  &= \mathbf{G}^{H} \mathbf{y}_n  \\
                &= \mathbf{G}^{H} \mathbf{H} \mathbf{x}_n + \mathbf{G}^{H}  \mathbf{z}_n \\
                &= \mathbf{H}^f \mathbf{a}_n + \mathbf{z}_n^f
  \end{aligned}
\end{equation}
where \( \mathbf{H}^f = \mathbf{G}^{H} \mathbf{H} \mathbf{G} \mathbf{\Phi}^H \) is the \( \Gamma \times \Gamma \) effective channel matrix after filtering and \( \mathbf{z}_n^f = \mathbf{G}^{H}  \mathbf{z}_n \).
Then, down-sampling followed by FFT is performed on the filtered signal to obtain the frequency domain signal vector \( \mathbf{r}_n  = [r_{0,n}, r_{1,n}, \ldots, r_{M-1,n}]\). The output of the FFT is given by
\begin{equation}
  \begin{aligned}
\label{EQ7}
\mathbf{r}_n &= \mathbf{\Phi} \mathbf{y}_n^f \\
             &= \mathbf{\Phi} \mathbf{H}^f \mathbf{a}_n + \mathbf{\Phi} \mathbf{z}_n^f \\
             &= \tilde{\mathbf{H}} \mathbf{a}_n + \tilde{\mathbf{z}}_n
  \end{aligned}
\end{equation}
where \( \tilde{\mathbf{H}} = \mathbf{\Phi} \mathbf{G}^{H} \mathbf{H} \mathbf{G} \mathbf{\Phi}^H \) represents the \( M \times M \) effective channel matrix after FFT and \( \tilde{\mathbf{z}} = \mathbf{\Phi} \mathbf{G}^{H}  \mathbf{z}_n \) is the coloured noise.
Then, the received FBMC-QAM signal associated with \(m\)-th subcarrier and the \(n\)-th symbol, \( r_{m,n}\), is expressed as
\begin{dmath}
\label{EQ8}
r_{m,n} = \tilde{\mathbf{H}}_{m,n} a_{m,n} + I_{m,n}^{intrinsic} + \tilde{z}_{m,n}
\end{dmath}
where \( I_{m,n}^{intrinsic} \) represents the intrinsic interference caused by the loss of complex orthogonality in FBMC-QAM, which is given by
\begin{dmath}
\label{EQ9}
I_{m,n}^{intrinsic} = \underbrace{\sum_{i \neq m} \tilde{\mathbf{H}}_{i,n}^{ICI} a_{i,n}}_{ICI}   + \underbrace{\sum_{j \neq n} \sum_{m = 0}^{M-1} \tilde{\mathbf{H}}_{m,j}^{ISI} a_{m,j}}_{ISI} 
\end{dmath}
where \(\tilde{\mathbf{H}}_{i,n}^{ICI} \) and \( \tilde{\mathbf{H}}_{m,j}^{ISI} \) are the residual channels that lead to inter-carrier interference (ICI) and inter-symbol interference (ISI), respectively. Note that the characteristics of \( \tilde{\mathbf{H}} \) is determined by the type of prototype filter and the fading channel effect. Therefore, by using a single prototype filter the effective channel will vary only with the channel fading effect unless a different prototype filter is selected.

As an illustration of the interference terms, consider the \(i\)-th transmitted symbol of a system with  \( M=5 \) subcarriers, i.e. \(\mathbf{x}^i = [x_0^i \:\: x_1^i \:\: x_2^i \:\: x_3^i \:\: x_4^i] \) and a channel with impulse response \(\mathbf{h}^i = [h_0 \:\: h_1 \:\: h_2 \:\: h_3 \:\: h_4] \). The effect of the channel on the transmitted symbol is shown in Fig. 3. Fig. 3 (a) shows the structure of ICI caused by other subcarriers on the same symbol whereas Fig. 3 (b) gives the ISI from one symbol to the other. For desirable performance, the ICI and ISI terms must be estimated and cancelled from the received signal. To do this, the interference channels \(\tilde{\mathbf{H}}_{i,n}^{ICI} \) and  \( \tilde{\mathbf{H}}_{m,j}^{ISI} \) must be estimated at the receiver\footnote{We assume that the receiver has knowledge of the channel, which can be obtained with channel estimation methods such as the one described in \cite{YahyaJasimHarbi:16}.}. The interference channel matrix can be expressed as \cite{YahyaJHarbi:16}
\begin{equation}
\label{EQ10}
\tilde{\mathbf{H}}_{i,j}^{I} = \begin{bmatrix}
    0_{E \times (N-E)} & \mathbf{H}_{E}  \\
    0_{(N-E) \times (N-E)} & 0_{(N-E) \times E}  \\
\end{bmatrix}
\end{equation} 
where \(I = ICI , ISI \),  \( \mathbf{H}_{E} \in \mathbb{C}^{E \times E} \), with \(E=L_{ch}-1\), is given by 
\begin{equation}
\label{EQ11}
\mathbf{H}_{E} = \begin{bmatrix}
    h_{L{ch}-1}       & \dots & \dots & h_{0} \\
    0       & \ddots &   & \vdots \\
    \vdots  & \ddots & \ddots & \vdots \\
    0       & \dots  &   0    & h_{L_{ch}-1}
\end{bmatrix}
\end{equation} 

\begin{figure}
  \begin{subfigure}{8cm}
    \centering\includegraphics[width=3.5in]{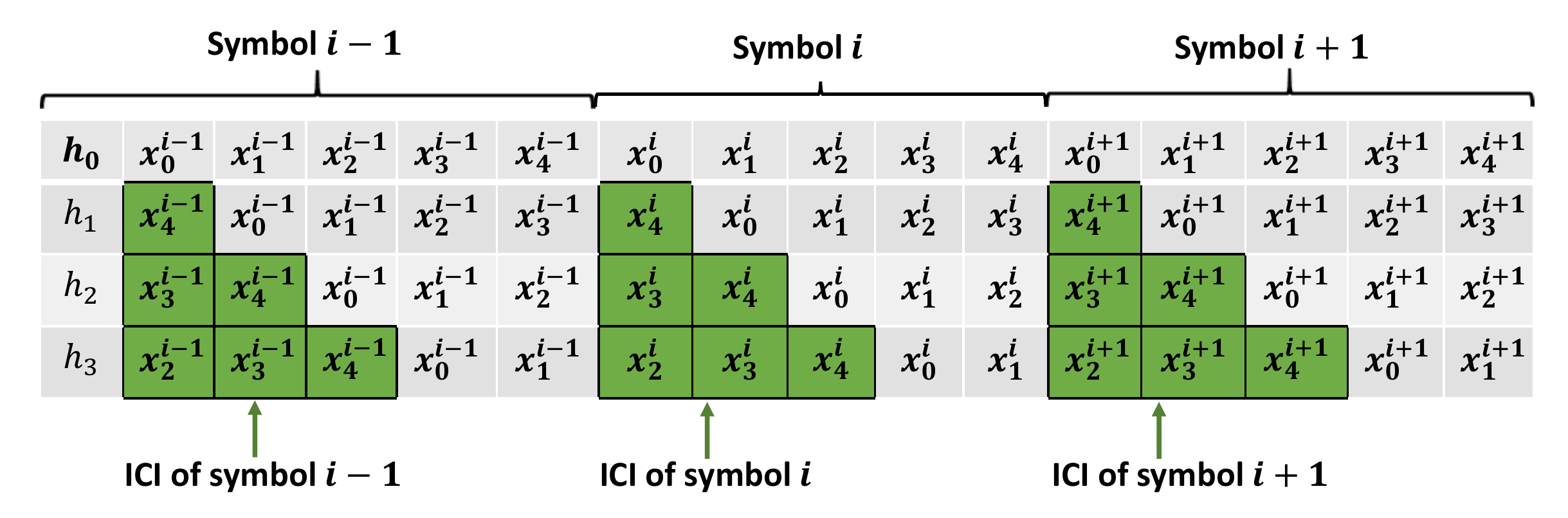}
    \caption{ICI}
  \end{subfigure}
  \begin{subfigure}{8cm}
    \centering\includegraphics[width=3.5in]{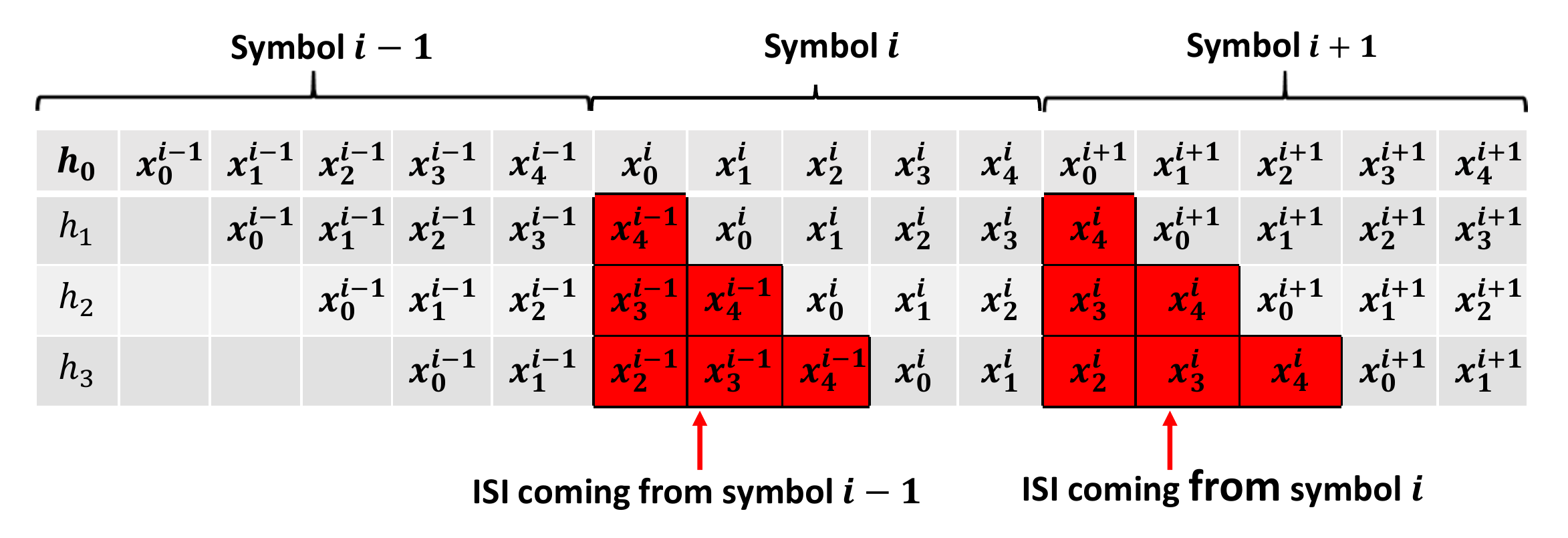}
    \caption{ISI}
  \end{subfigure}
\caption{Effect of frequency-selective channel on received FBMC-QAM symbols} 
\end{figure}

Before detection and decoding of the received signal, the effects of the frequency-selective channel is compensated by applying a simple one-tap zero-forcing (ZF) equalizer. The resulting signal is represented as 
\begin{equation}
  \begin{aligned}
\label{EQ12}
\tilde{r}_{m,n} &= \frac{r_{m,n}}{\tilde{\mathbf{H}}_{m,n}} \\
                &= a_{m,n} + \frac{I_{m,n}^{intrinsic}}{\tilde{\mathbf{H}}_{m,n}} + \frac{\tilde{z}_{m,n}}{\tilde{\mathbf{H}}_{m,n}}
  \end{aligned}
\end{equation}
The equalized signal is deinterleaved and passed to the LDPC decoder for decoding and detection. The receiver processing is repeated for several iterations to get rid of the intrinsic interference through decoding and IIC. This is described in detail in the next subsection. 

\begin{figure*}
\centering
\includegraphics[width=6in]{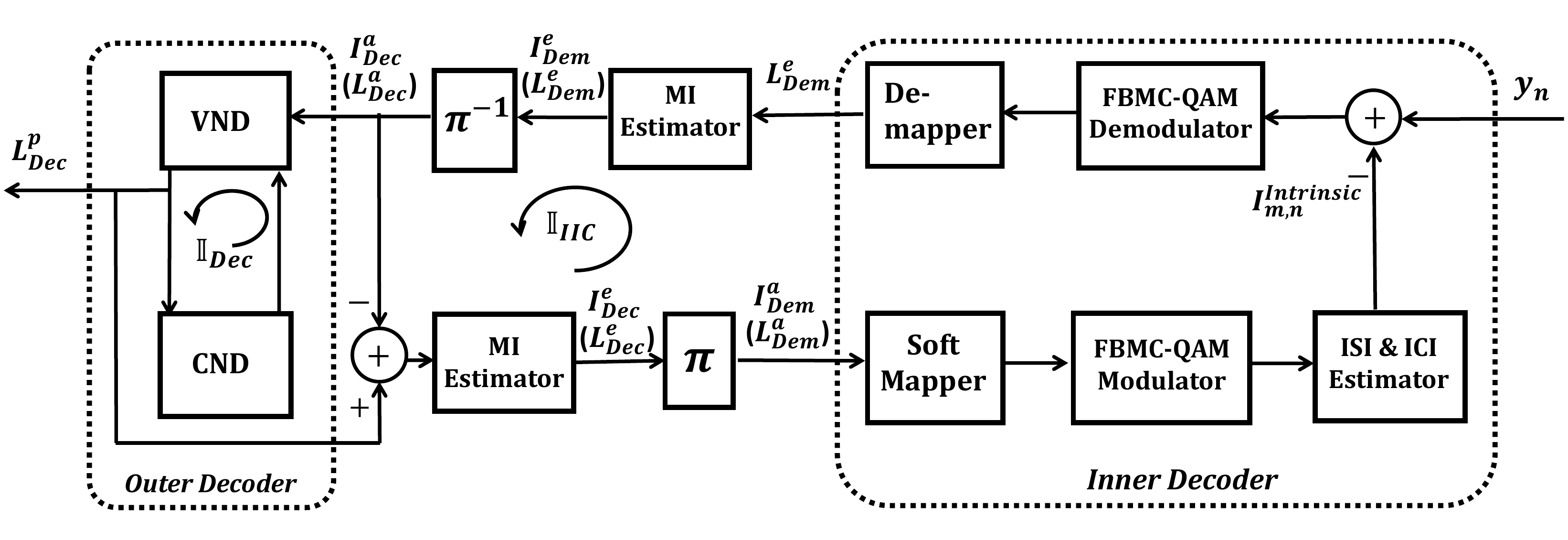}
\caption{IIC-based BICM-ID system model for EXIT Chart analysis}
\end{figure*}

\subsection{IIC-based BICM-ID Receiver}
In order to recover the transmitted bits, an iterative detection and decoding receiver is proposed as shown in Fig. 4 on the next page. It is made up of two component decoders:
\begin{itemize}
\item The \textit{Inner Decoder} - which consists of the soft mapper, soft demapper, FBMC-QAM modulator and demodulator, and an iterative interference cancellation (IIC) operation.
\item The \textit{Outer Decoder} - representing the LDPC decoder, which consists of two types of nodes: VND and CND.
\end{itemize}
To remove the intrinsic interference caused by the loss of orthogonality in FBMC-QAM, the proposed BICM-ID receiver performs two iterative processes: (i) the exchange of mutual information (MI) between the VND and CND of the outer decoder and (ii) the exchange of MI between the inner decoder and outer decoder. The inner decoder takes the received signal \( y_n \) and the \textit{a priori} information of the coded bits, \( L_{Dem,a}^{q,n} \), from the outer decoder (\( L_{Dem,a}^{q,n} = 0 \) in the first receiver iteration) and computes the \textit{a posteriori} log-likelihood ratios (LLRs) as 
\begin{equation}
\label{EQ13}
L_{Dem}^{q,n} = L_{Dem,a}^{q,n} + L_{Dem,e}^{q,n}
\end{equation}
where \( L_{Dem,e}^{q,n} \) is the extrinsic LLR values of the inner decoder. Using the maximum  \textit{a posteriori} demapping algorithm, \( L_{Dem,a}^{q,n} \) and \( L_{Dem,e}^{q,n} \) can be expressed as \cite{QLi:16}
\begin{equation}
\label{EQ14}
L_{Dem,a}^{q,n} = \log \frac{p(a_{n}^{\mathit{q}} = +1 \mid \tilde{r}_{n})}{p(a_{n}^{\mathit{q}} = -1 \mid \tilde{r}_{n})}
\end{equation}
and 
\begin{equation}
\label{EQ15}
L_{Dem,e}^{q,n} = \log \frac{\sum\limits_{a \in a_q^+} p(\tilde{r}_{n} \mid a) \prod\limits_{\substack{q' = 1 \\ q' \neq q}} p(a_{n}^{\mathit{q'}})}{\sum\limits_{a \in a_q^-} p(\tilde{r}_{n} \mid a) \prod\limits_{\substack{q' = 1 \\ q' \neq q}} p(a_{n}^{\mathit{q'}})}
\end{equation}

\begin{algorithm}[t]
\caption{IIC-based BICM-ID Receiver Algorithm}
\label{alg:MGSAlgorithmWPCN}
\begin{algorithmic}[1] 
\STATE \textbf{Require}: \( \mathbf{r}_n \), \( \mathbb{I}_{Dec}^{max}\), \( \mathbb{I}_{IIC}^{max}\)  
\STATE Initialize number of iterations \( i = 0 \)
\WHILE{\( i \leq \mathbb{I}_{IIC}^{max} \)} 
\STATE Perform FBMC-QAM demodulation of received signal.
\STATE Pass output of FBMC-QAM demodulator to soft demapper to calculate extrinsic LLR values.
\STATE Deinterleave extrinsic LLR values to obtain \textit{a priori} LLR for the LDPC decoder, \( L_{Dec,a}^{q,n} \).
\WHILE{\( i \leq \mathbb{I}_{Dec}^{max} \)} 
\STATE Pass \( L_{Dec,a}^{q,n} \) between VND and CND for defined number of iterations.
\ENDWHILE
\STATE Interleave extrinsic LLR values of LDPC decoder to obtain \textit{a priori} LLR for soft mapper.
\STATE Determine estimate of the transmit signal \( \hat{\mathbf{a}}_n \), using a soft MQAM mapper. 
\STATE Pass the estimated signal through the FBMC-QAM modulator.
\STATE Estimate intrinsic interference as shown in (\ref{EQ9}).
\STATE Subtract the estimated interference terms from the received signal.  
\STATE \( i = i + 1 \)
\ENDWHILE
\STATE Make hard decision to obtain final estimates of the transmitted bits, \( \hat{\mathbf{b}} \).
\STATE Terminate.
\end{algorithmic}
\end{algorithm}

The extrinsic LLR values, \( L_{Dem,e}^{q,n} \), are de-interleaved and passed to the outer decoder as \textit{a priori} LLR, \( L_{Dec,a}^{q,n} \), for channel decoding. After a number of iterations between the CND and VND of the outer decoder, it computes \textit{a posteriori} LLR, \( L_{Dec,p}^{q,n} \). The outer decoder extrinsic information, defined as \( L_{Dec,e}^{q,n} = L_{Dec,p}^{q,n} - L_{Dec,a}^{q,n}\), is re-interleaved and passed to the soft mapper as \textit{a priori} information, \(L_{Dem,a}^{q,n}\). To estimate and cancel the effect of the intrinsic interference term in (\ref{EQ12}), the output of the soft mapper is FBMC-QAM modulated and the estimated symbols are subtracted from the received signal for the next iteration as shown in Fig. 4. We denote the number of iterations between the inner decoder and the outer decoder by \( \mathbb{I}_{IIC} \) and the number of iterations within the outer decoder by \( \mathbb{I}_{Dec} \). After the final iteration in both component decoders, \( L_{Dec,p}^{q,n} \) is used to generate the hard-decision estimates of the transmitted bits. The iterative process of the proposed IIC-based BICM-ID receiver proceeds as shown in Algorithm 1 on the previous page.

\section{Convergence Analysis of IIC-Based BICM-ID Receiver}
EXIT charts are useful for studying the convergence of iterative decoders \cite{StenBrink:04}. The EXIT chart analysis tracks the exchange of MI between the components of an iterative receiver in order to predict the required number of iterations and convergence threshold of the receiver. In this section, we derive the MI for the inner and outer decoders and study their evolution using EXIT charts in order to visualize the convergence behaviour of the proposed IIC-based BICM-ID receiver. 

\subsection{Mutual Information}
The MI between the transmitted bits and the LLR values at the output of a component decoder can be estimated by assuming a known probability distribution for the LLR values. The MI between equally likely information bits, \(x\) and their corresponding LLR values, \(L\), can be expressed as \cite{JHagenauer:04}
\begin{equation}
\begin{aligned}
\label{EQ16}
I(x,L) &= \frac{1}{2} \sum\limits_{x = \pm 1} \int_{-\infty}^{+\infty} 
p_L (\lambda | X=x) \\ 
         & \times \log_2 \frac{p_L (\lambda | X=x)}{p_L(\lambda)} d \lambda
\end{aligned}
\end{equation}
with
\begin{equation}
\label{EQ17}
p_L(\lambda) = \frac{1}{2} \left[ \: p_L (\lambda | X=-1) + p_L (\lambda | X=+1) \: \right]
\end{equation}
where \( p_L (\lambda | X=x) \) is the conditional probability density function (PDF) associated with \( L \) and \( X \) is a random variable (RV) representing the information bits. The LLR values, \( L \), can be modelled by an independent Gaussian RV \( n_L \) with zero mean and variance \( \sigma_L^2 \) given by
\begin{equation}
\label{EQ18}
L = \mu_L x + n_L
\end{equation}
where \( \mu_L = \sigma_L^2/2\) and \( \sigma_L^2 \) can be computed as \cite{Brannstrom:05}
\begin{equation}
\label{EQ19}
\sigma_L^2 \approx \left( - \frac{1}{H_1} \log_2 \left( 1 - I(x,L)^{1/H_3} \right) \right)^{1/2H_2}
\end{equation}
where \(H_1 = 0.3073\), \(H_2 = 0.8935\) and \(H_3 = 1.1064\).
Following the Gaussian PDF, the MI expression in (\ref{EQ16}) can be rewritten as 
\begin{equation}
\begin{aligned}
\label{EQ20}
I_L (\sigma_L) &= 1 - \frac{1}{\sqrt{2 \pi} \sigma_L} \int_{-\infty}^{+\infty} \exp \left(- \frac{ \left(  \lambda-\frac{\sigma_L^2}{2} \right)^2}{2 \sigma_L^2} \right) \\
                 & \times \log_2 \left[ \: 1+ \exp(-\lambda) \: \right] d \lambda. 
\end{aligned}
\end{equation}
Assuming a symmetric PDF for the LLR values, i.e. \( p_L (-\lambda | X=+1) =  p_L (\lambda | X=-1) \), a simple approximation of the MI expression in (\ref{EQ20}) is given by 
\begin{equation}
\label{EQ21}
\begin{aligned}
I(x,L) = 1 - \mathbb{E} \{\log_2 \left[ \: 1+ \exp(-\lambda) \: \right]\}
\end{aligned}
\end{equation}
where \( \mathbb{E}  \) represents the expectation operation. According to \cite{JHagenauer:04}, by replacing the ensemble average in (\ref{EQ21}) by the time average, the MI can be measured for a sufficiently large number of samples, \( T \), even for non-Gaussian or undefined distributions\footnote{The proof for MI follows from \cite{JHagenauer:04} and is therefore omitted in this paper for brevity.}. The approximated MI expression using time averaging is given as
\begin{equation}
\label{EQ22}
I(x,L) = 1 - \frac{1}{T} \sum\limits_{t = 1}^{T} \log_2 [ \: 1 + \exp(-x_t \cdot L_t )\: ]
\end{equation}
where \(L_t\) is the LLR value associated with the bits \(x_t \in \{-1,+1\}\).

\subsection{EXIT Chart of Inner Decoder}
Notice from Fig. 4 that the MI output of the inner decoder is dependent on two inputs, namely, the \textit{a priori} MI, \(I_{Dem}^{a}\), coming as feedback from the outer decoder and the received signal coming from the channel. Thus, the inner decoder's EXIT characteristics can be defined by the \textit{a priori}-extrinsic transfer function \(\mathbb{T}\) as 
\begin{equation}
\label{EQ23}
I_{Dem}^{e} = \mathbb{T}[ \: I_{Dem}^{a},\sigma_n^2 \: ]
\end{equation}
where  \(I_{Dem}^{e} = I[x;L_{e,Dem}^{q,n}], \: 0 \leq I_{Dem}^{e} \leq 1 \), is the extrinsic MI of the inner decoder and \( \sigma_n^2 \) is the channel noise variance. \(I_{Dem}^{e} \) can be obtained from (\ref{EQ22}) as
\begin{equation}
\label{EQ24}
I_{Dem}^{e} = 1 - \frac{1}{T} \sum\limits_{t = 1}^{T} \log_2 [1 + \exp(-x_t \cdot
L_{e,Dem}^{q,n} )]
\end{equation}
The extrinsic LLR of the inner decoder, \(L_{e,Dem}^{q,n}\), is then deinterleaved and passed to the outer decoder as \textit{a priori} LLR, \( L_{a,Dec}^{q,n} \).

\subsection{EXIT Chart of Outer Decoder}
The outer decoder accepts \textit{a priori} input, \( L_{a,Dec}^{q,n} \), from the inner decoder and alternates this information between the VND and CND processing for a predefined number of iterations, \(\mathbb{I}_{Dec} \). When the decoding process is complete it returns extrinsic LLR, \( L_{e,Dec}^{q,n} \). This is represented by the transfer function 
\begin{equation}
\label{EQ25}
I_{Dec}^{e} = \mathbb{T}[ \: I_{Dec}^{a} \: ]
\end{equation}
where \(I_{Dec}^{a} = I[x;L_{a,Dec}^{q,n}], \: 0 \leq I_{Dec}^{a} \leq 1 \), is the MI between the encoded bits stream and the \textit{a priori} LLR, \( L_{a,Dec}^{q,n} \) and \(I_{Dec}^{e} = I[x;L_{e,Dec}^{q,n}], \: 0 \leq I_{Dec}^{e} \leq 1 \) is the extrinsic MI of the outer decoder which can be computed according to (\ref{EQ22}) as
\begin{equation}
\label{EQ26}
I_{Dec}^{e} = 1 - \frac{1}{T} \sum\limits_{t = 1}^{T} \log_2 [1 + \exp(-x_t \cdot
L_{e,Dec}^{q,n} )]
\end{equation}
For the next IIC iteration the extrinsic LLR from the outer decoder, \(L_{e,Dec}^{q,n}\) is re-interleaved and passed to the inner decoder as \textit{a priori} LLR \( L_{a,Dem}^{q,n} \). 

\subsection{Convergence Analysis}
Given the MI of the inner and outer decoders, the convergence behaviour of the IIC-based BICM-ID receiver can be visualized by plotting the EXIT characteristics of the two decoders on an EXIT chart. The inner decoder EXIT curve is dependent on the noise variance or the signal-to-noise ratio (SNR) value of the received signal while the outer decoder EXIT curve varies with the number of decoder iterations \( \mathbb{I}_{Dec} \). The outer decoder's extrinsic MI, \(I_{Dec}^{e} \), is feedback to the inner decoder as \textit{a priori} MI \( I_{Dem}^{a} \) and is plotted on the x-axis of the EXIT chart. Likewise, the inner decoder's extrinsic MI \(I_{Dem}^{e} \) is passed to the outer decoder as \textit{a priori} information \( I_{Dec}^{a} \), which represents the y-axis of the EXIT chart. For near capacity decoding performance, a pair of inner and outer decoder EXIT curves should not intersect before the (1,1) point on the EXIT chart. In this case, an open area or tunnel exists between the two curves and the exchange of MI can be visualized as a "zig-zag" trajectory on the EXIT chart as shown in Fig. 6(a). The wider the tunnel, the lower the number of iterations required to reach the (1,1) point and vice versa. On the other hand, if any pair of inner and outer decoder EXIT curves intersect before reaching  \( I_{Dem}^{a} = 1 \) on the horizontal axis, the tunnel is said to be "blocked". In this case the EXIT chart does not converge and a poor BER performance is obtained even with a high number of IIC iterations. Therefore, to achieve near capacity performance at a low receiver complexity, a trade-off between the SNR values, number of outer decoder iterations and the width of the EXIT tunnel must be obtained. The convergence behaviour of the proposed receiver is studied by Monte-Carlo simulations in the next section. 

\section{Performance and Complexity Evaluation}
In this section, numerical simulation results are presented to illustrate the OOB emission performance of different FBMC systems compared to CP-OFDM. Also, the EXIT charts and BER performance of the proposed IIC-based BICM-ID receiver under different fading channel conditions are shown. The complexity of the proposed receiver for FBMC-QAM is studied and compared to the complexity of a CP-OFDM benchmark. First, we introduce the simulation parameters. 

\begin{table}[t!]
\label{Tab:1}
\caption{Simulation Parameters}
\begin{center} \vline
     \begin{tabular}{ l | l }
     \hline 
      \textbf{Parameter } & \textbf{Specification}  \\ \hline
      Filter & PHYDYAS prototype filter \cite{Bellanger:10} \\ \hline
      Overlapping factor (\(K\)) & 4 \\ \hline
      Channel Bandwidth & 1.4 MHz  \\ \hline
      Total number of subcarriers & 12  \\ \hline
      Number of resource blocks (RBs) & 6  \\ \hline
      Number of subcarriers per RBs & 12  \\ \hline
      Subcarriers spacing & 15 KHz  \\ \hline
      Number of slots per RBs & 2  \\ \hline
      Number of symbols per slot & 7  \\ \hline
      Modulation  & 16-QAM, \\ \hline
      LDPC Code rate  & 1/2 \\ \hline
      Mapping     & \( M 16^r \) \\ \hline
      Number of IIC iterations  & 0, 1, 2, 3 \\ \hline
      Channel models & LTE-EPA, LTE-EVA, LTE-ETU  \\ \hline 
     \end{tabular} 
\end{center} 
\end{table}

\begin{table}[t!]
\label{Tab:2}
\caption{LTE Channel Delay and Power Profile}
\begin{center} \vline
   \begin{tabular}{|c|c|c|c|c|c|}
   \hline
   \multicolumn{3}{c|}{Delay (\textit{ns})}  & 
   
    \multicolumn{3}{c|}{Power (\textit{dB})}  \\ \hline

       EPA  & EVA  & ETU  & EPA   & EVA   & ETU \\ \hline
       0    & 0    & 0    &  0    & 0     & -1.0 \\ \hline
       30   & 30   & 50   & -1.0  & -1.5  & -1.0  \\ \hline
       70   & 150  & 120  & -2.0  & -1.4  & -1.0 \\ \hline
       90   & 310  & 200  & -3.0  & -3.6  & 0 \\ \hline
       110  & 370  & 230  & -8.0  & -0.6  & 0 \\ \hline
       190  & 710  & 500  & -17.2 & -9.1  & 0 \\ \hline
       410  & 1090 & 1600 & -20.8 & -7.0  & -3.0 \\ \hline
            & 1730 & 2300 &       & -12.0 & -5.0 \\ \hline
            & 2510 & 5000 &       & -16.9 & -7.0 \\ \hline
  \end{tabular} 
\end{center} 
\end{table}

\subsection{Simulation Setup}
\label{Simulation:Setup}
To evaluate the effectiveness of the IIC-based BICM-ID receiver for FBMC-QAM, computer simulations are conducted. For the LDPC decoder, an irregular parity-check matrix has been used whereas the soft decision message-passing algorithm, sum-product decoding, is employed as demapper \cite{SarahJohnson:09}. To achieve effective convergence in the EXIT chart analysis, anti-gray mapping schemes have been proposed in the literature. Scheme such as set partitioning (SP), modified set partitioning (MSP), maximum squared Euclidean weight (MSEW) and the optimized mapping \( M 16^r \) have been shown to perform better than classical Gray mapping in EXIT chart analysis \cite{Schreckenbach:03}. Therefore, \( M 16^r \) is adopted in this simulation. For comparison, synchronous CP-OFDM is considered as a benchmark. For the benchmark implementation, we assume that the CP-OFDM system has sufficient CP and guard band in order to maintain orthogonality and synchronization between subbands. We also set the number of outer decoder  iterations for CP-OFDM to \( \mathbb{I}_{Dec} = 8 \). This is because, as shown in Fig. 6(a), the MI of the outer decoder converges after 8 iterations, with negligible difference in MI from 8 to 10 iterations.  Note that an FBMC-QAM system with a single prototype filter for all subcarriers is considered in this paper. For the filter implementation, the PHYDYAS prototype filter with overlapping factor \( K = 4\) is used. The prototype filter coefficients are defined as \cite{Chen:15}
\begin{equation}
\label{EQ27}
\begin{aligned}
g[i] &= P(0) + 2 \sum_{k=1}^K (-1)^k P(k) \cos \left( \frac{2 \pi k i}{\Gamma} \right)
\end{aligned}
\end{equation}
where \( P(0) = 1 \), \( P(1) = 0.971960 \), \( P(2) = 1/\sqrt{2} \) and  \( P(3) = 0.235147 \). As mentioned above, the choice of filter is motivated by results in \cite{Zakaria:14} which shows that the PHYDYAS filter achieves good performance when applied to interference cancellation receivers. Also, PSD of different prototype filters are compared in \cite{Chen:18} and the results show that the PHYDYAS filter has the fastest sidelobe attenuation speed (i.e. low OOB emission).

To study the effect of fading on the proposed system, we consider the 3GPP standardized channel models Extended Pedestrian A (EPA), Extended Vehicular A (EVA) and Extended Typical Urban (ETU). Perfect channel state information is assumed at both the transmitter and receiver. The simulation parameters and fading channel profiles are presented in Table I and II, respectively.

\subsection{OOB Emission Performance} 
\label{SubSecResults3}
The power spectral density (PSD) of CP-OFDM, FBMC-QAM with twin filters and FBMC-QAM with a single filter are given in Fig. 5. The FBMC-QAM system with twin filters, presented in \cite{HNam:16}, adopts the conventional PHYDYAS filter for even-numbered subcarriers and a reordered version of it for odd-numbered subcarriers in order to achieve orthogonality in AWGN channels. As can be seen, it achieves OOB emission that is worse than that of CP-OFDM. In frequency selective channels the orthogonality is lost because of the presence of residual interference, which degrades the BER performance. The optimized twin filters in \cite{Yun:15} are designed to find a trade-off between intrinsic interference and OOB performance. As shown in Fig. 5, the optimized filter set improves the OOB performance compared to both CP-OFDM and the filter set in \cite{HNam:16}. The in-band spectrum of the different waveforms are also shown in Fig. 5. The twin-filter waveforms show an in-band fluctuation, as was the case in \cite{Jintae:18}, which is absent in both CP-OFDM and FBMC-QAM with single filter. On the other hand, by completely relaxing the orthogonality condition it can be seen that FBMC-QAM with single prototype filter has much lower OOB leakage than CP-OFDM and FBMC-QAM with twin filters. The ultra-low OOB emission means there is very low leakage interference between subbands (or users occupying these subbands). This implies that in asynchronous communications, FBMC-QAM users occupying adjacent subbands will cause negligible interference to each other. For mMTC applications, synchronous communication will be extremely difficult to manage due to the large number of user terminals. Moreover, the signaling overhead associated with the synchronous communication in CP-OFDM can consume a significant amount of the available time and frequency resources. Therefore, FBMC-QAM can enable the benefits of asynchronous user transmissions to improve system capacity in various mMTC applications. An analysis of inter-user interference in a multi-user system to show the advantage of FBMC-QAM over CP-OFDM in asynchronous applications will be the topic of future work, but is beyond the scope of the present paper.   

\begin{figure}[t!]
\centering
\includegraphics[width=3.5in]{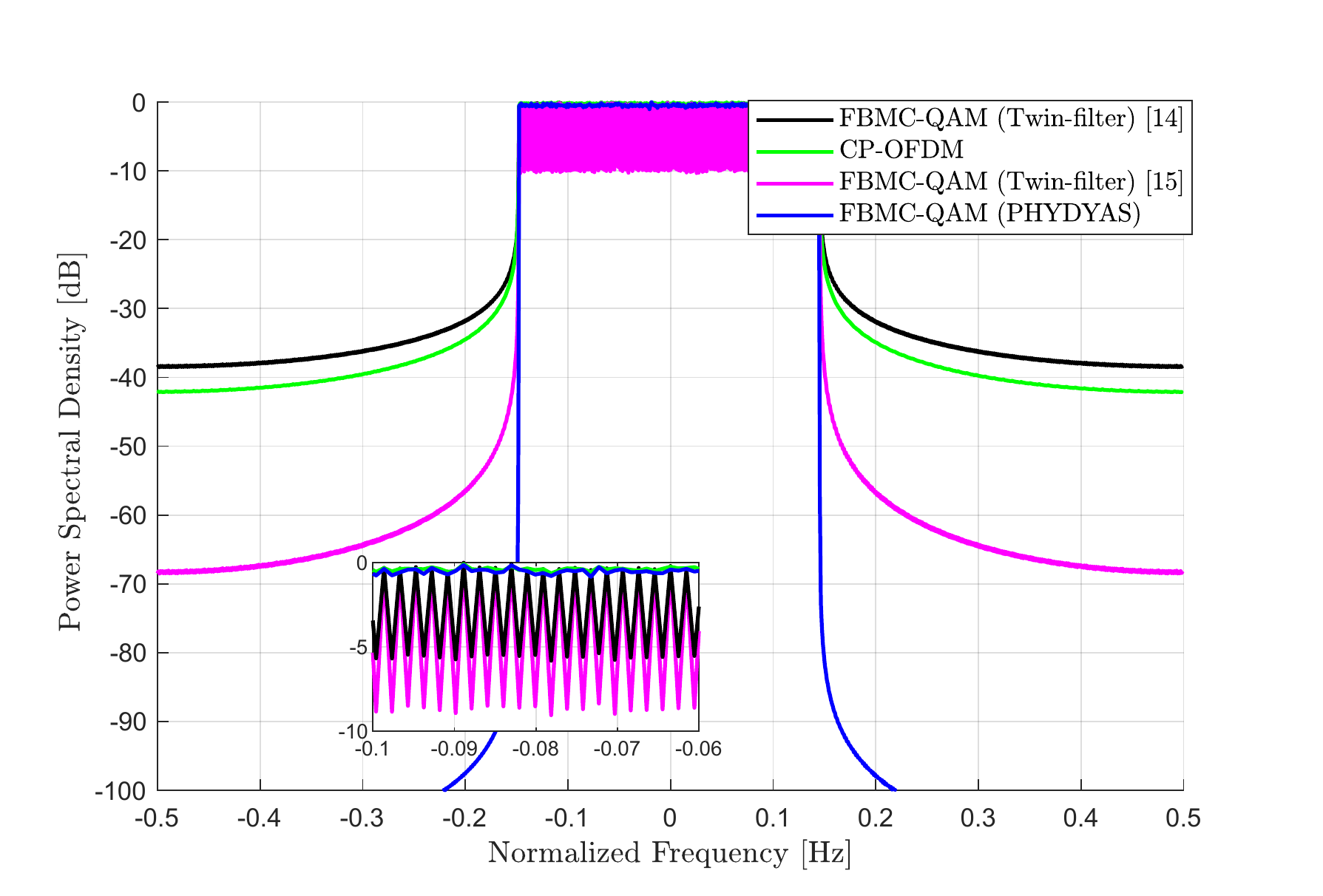}
\caption{PSD comparison of CP-OFDM, FBMC-QAM with twin filters and FBMC-QAM with single filter.}
\end{figure}

\begin{figure}
  \begin{subfigure}{6cm}
    \centering\includegraphics[width=3.3in]{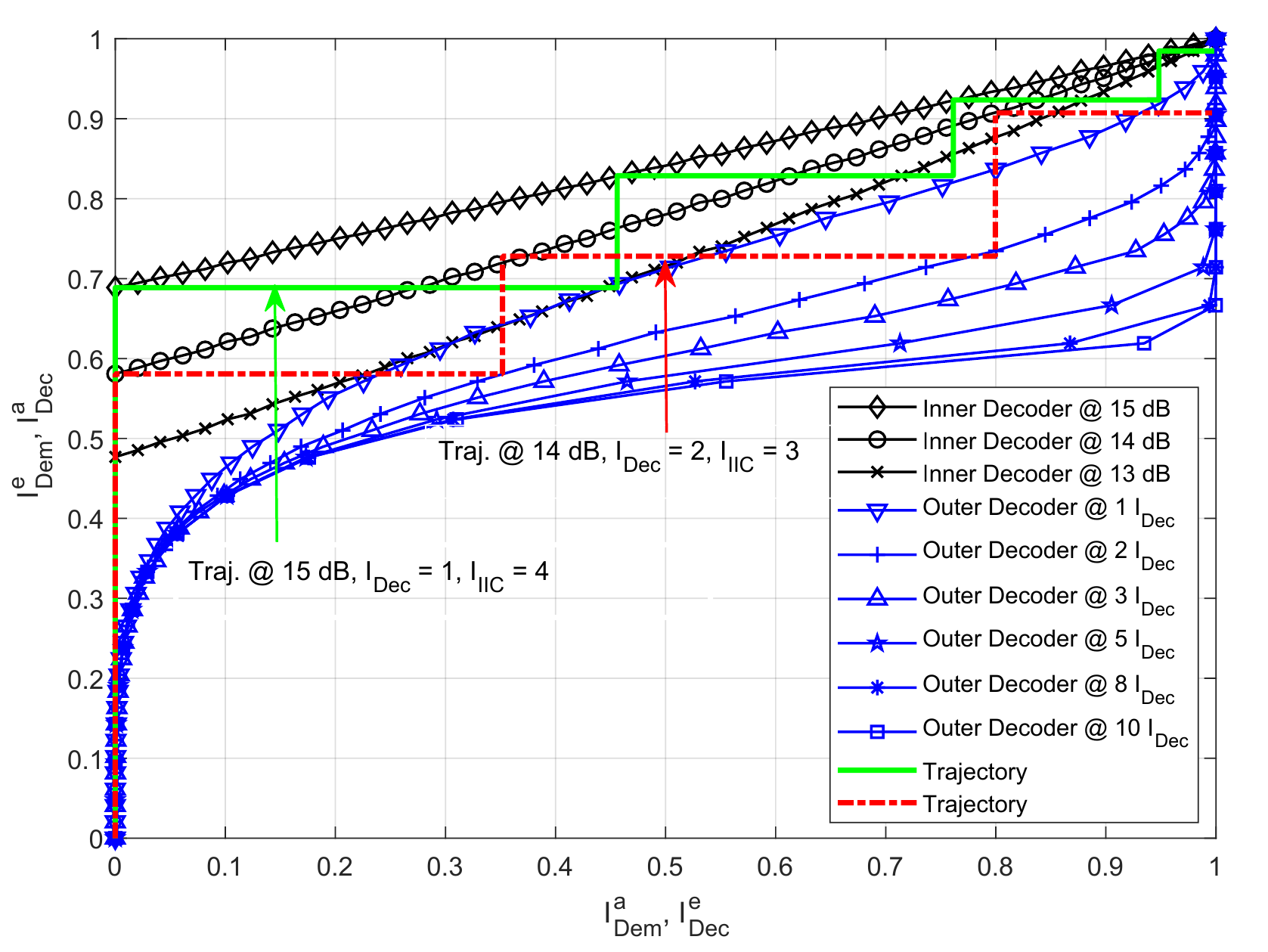}
    \caption{EXIT function and decoding trajectories}
  \end{subfigure}
  \begin{subfigure}{6cm}
    \centering\includegraphics[width=3.3in]{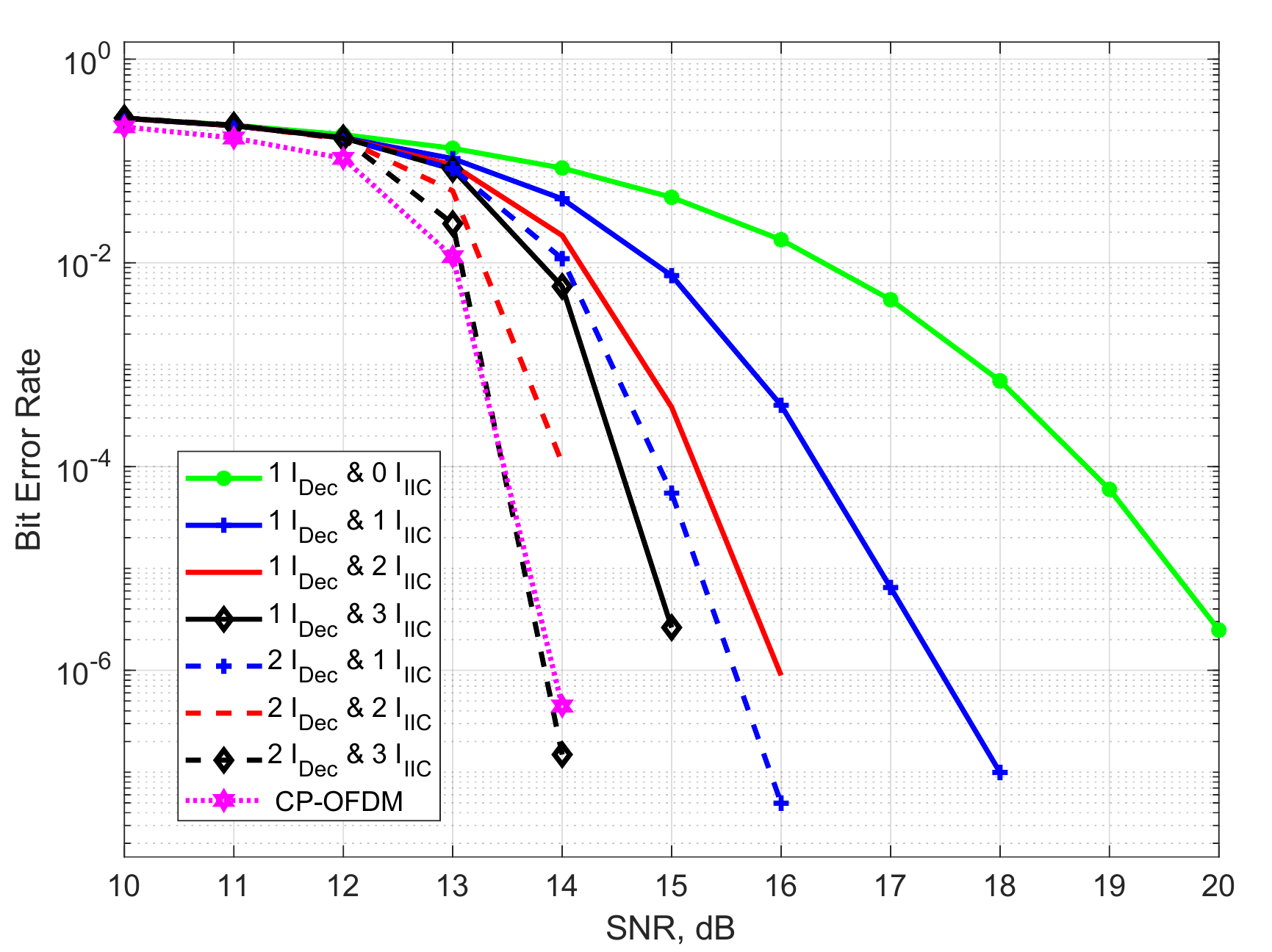}
    \caption{BER performance proposed receiver for FBMC-QAM and CP-OFDM}
  \end{subfigure}
\caption{Performance of the proposed IIC-based BICM-ID system with M\(16^r\) mapped 16-QAM over EPA channel} 
\end{figure}  

\begin{figure}
  \begin{subfigure}{6cm}
    \centering\includegraphics[width=3.3in]{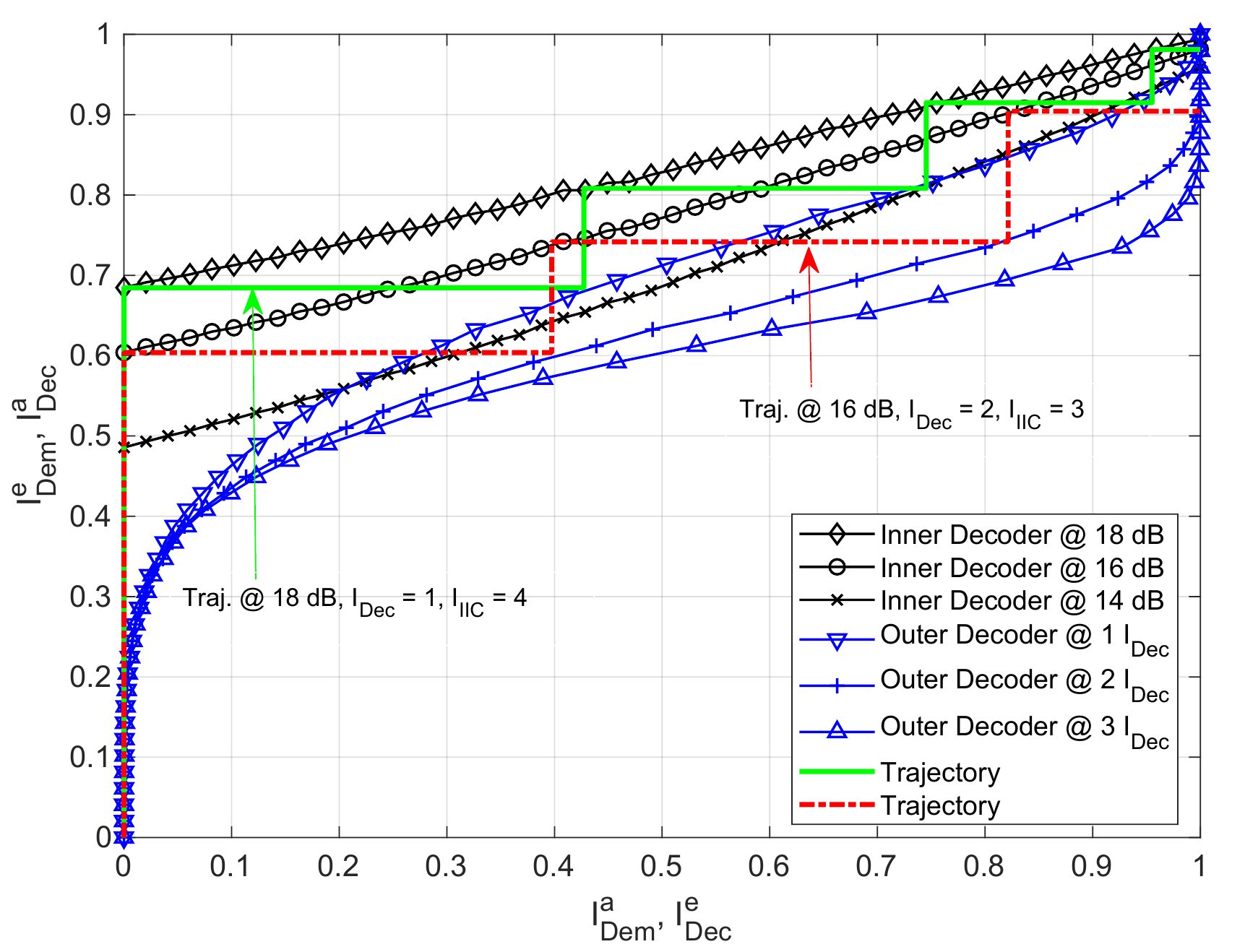}
    \caption{EXIT function and decoding trajectories}
  \end{subfigure}
  \begin{subfigure}{6cm}
    \centering\includegraphics[width=3.3in]{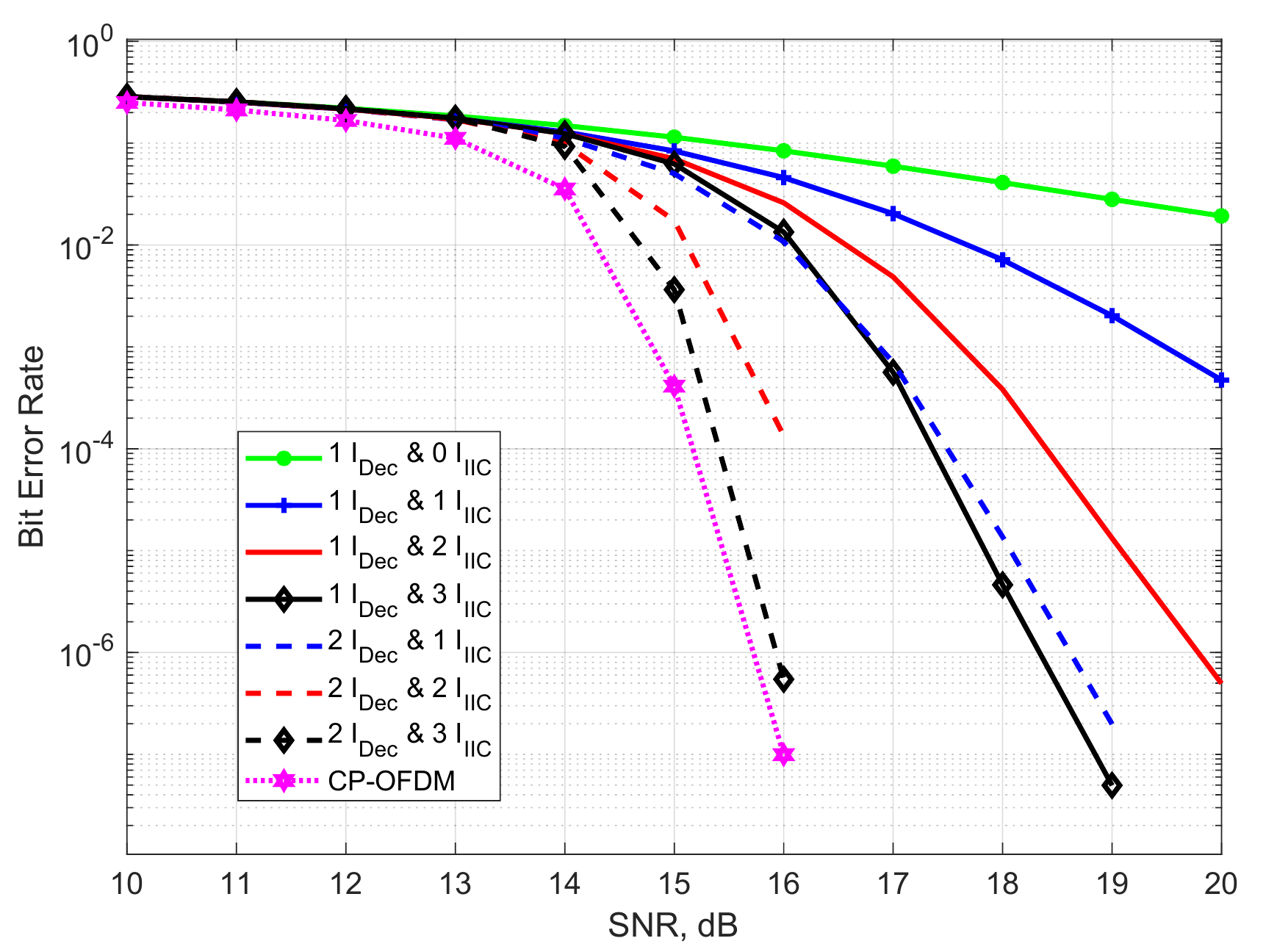}
    \caption{BER performance proposed receiver for FBMC-QAM and CP-OFDM}
  \end{subfigure}
\caption{Performance of the proposed IIC-based BICM-ID system with M\(16^r\) mapped 16-QAM over EVA channel} 
\end{figure}

\subsection{Performance of Single Fading Channel Realization}
\label{SubSecResults1}
Fig. 6 shows the performance of the proposed IIC-based BICM-ID receiver for FBMC-QAM and synchronous CP-OFDM over a single realization of the LTE EPA channel. Note that the MI of the inner decoder depends on the SNR values while the MI of the outer decoder varies with the number of iterations, \(\mathbb{I}_{Dec}\). The EXIT curves of the inner and outer decoders are shown in Fig. 6(a), with the "zig-zag" trajectories representing the exchange of MI between the inner decoder and the outer decoder. Observe from Fig. 6(a) that at SNR of 13dB and \(\mathbb{I}_{Dec} = 1\), the EXIT curve of the two component decoders intersect at \( I_{Dem}^{a} \approx 0.25 \). Therefore, there is no open tunnel between the two curves and the decoding iteration does not converge. In this case, the tunnel between the two EXIT curves is said to be blocked and increasing the number of IIC iterations, \(\mathbb{I}_{IIC}\), will not improve the system performance. However, by increasing either the SNR or \(\mathbb{I}_{Dec}\), the area between the two curves widens, resulting in an improved convergence behaviour of the BICM-ID receiver. For example, by increasing the SNR to 14dB and setting \(\mathbb{I}_{Dec} = 2\) a tunnel opens between the two EXIT curves. The number of trajectories between the two curves show the number of iterations required to achieve convergence. The wider the tunnel between the two curves, the lower the number of trajectories (IIC iterations) required to reach the (1,1) point on the EXIT chart and the faster the receiver converges. The information on the number of iterations (\(\mathbb{I}_{Dec}\) and \(\mathbb{I}_{IIC}\)) provided by the EXIT chart analysis can enable the prediction of the complexity of the proposed BICM-ID receiver. 

The results on the EXIT chart is validated by the BER performance presented in Fig. 6(b). As can be seen, for \(\mathbb{I}_{Dec} = 1\) the convergence threshold (turbo cliff/waterfall region) is obtained around 14dB. Therefore, by increasing the number of IIC iterations the BER performance is significantly improved. Measured at a BER of \(10^{-6}\) for \(\mathbb{I}_{IIC} \) values of 1, 2 and 3, the proposed receiver show about 3dB, 4dB and 5dB SNR gain, respectively, when compared with a receiver with no IIC iterations. Moreover, by increasing \(\mathbb{I}_{Dec}\) from 1 to 2 the convergence threshold occurs below 14dB. Hence, the decoding performance can be further improved with increasing IIC iterations. Compared to the benchmark CP-OFDM system, FBMC-QAM show about 6dB loss at \(10^{-6}\) BER when no IIC iterations are applied. Setting \(\mathbb{I}_{Dec} = 2 \) and \(\mathbb{I}_{IIC} = 3\), FBMC-QAM achieve the same BER performance as the CP-OFDM benchmark. However, increasing the iterations leads to increased receiver complexity. 

\begin{figure}
  \begin{subfigure}{6cm}
    \centering\includegraphics[width=3.3in]{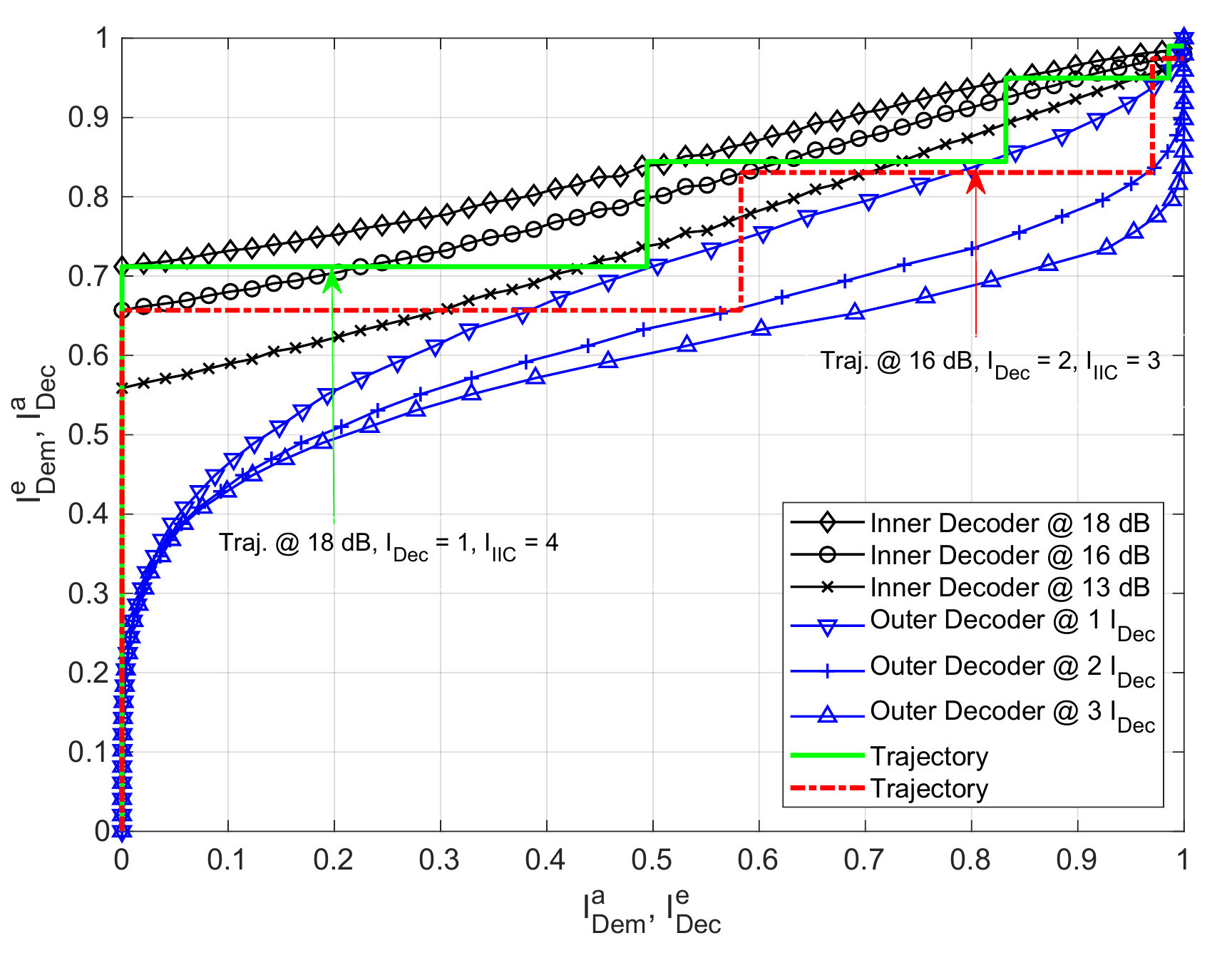}
    \caption{EXIT function and decoding trajectories}
  \end{subfigure}
  \begin{subfigure}{6cm}
    \centering\includegraphics[width=3.3in]{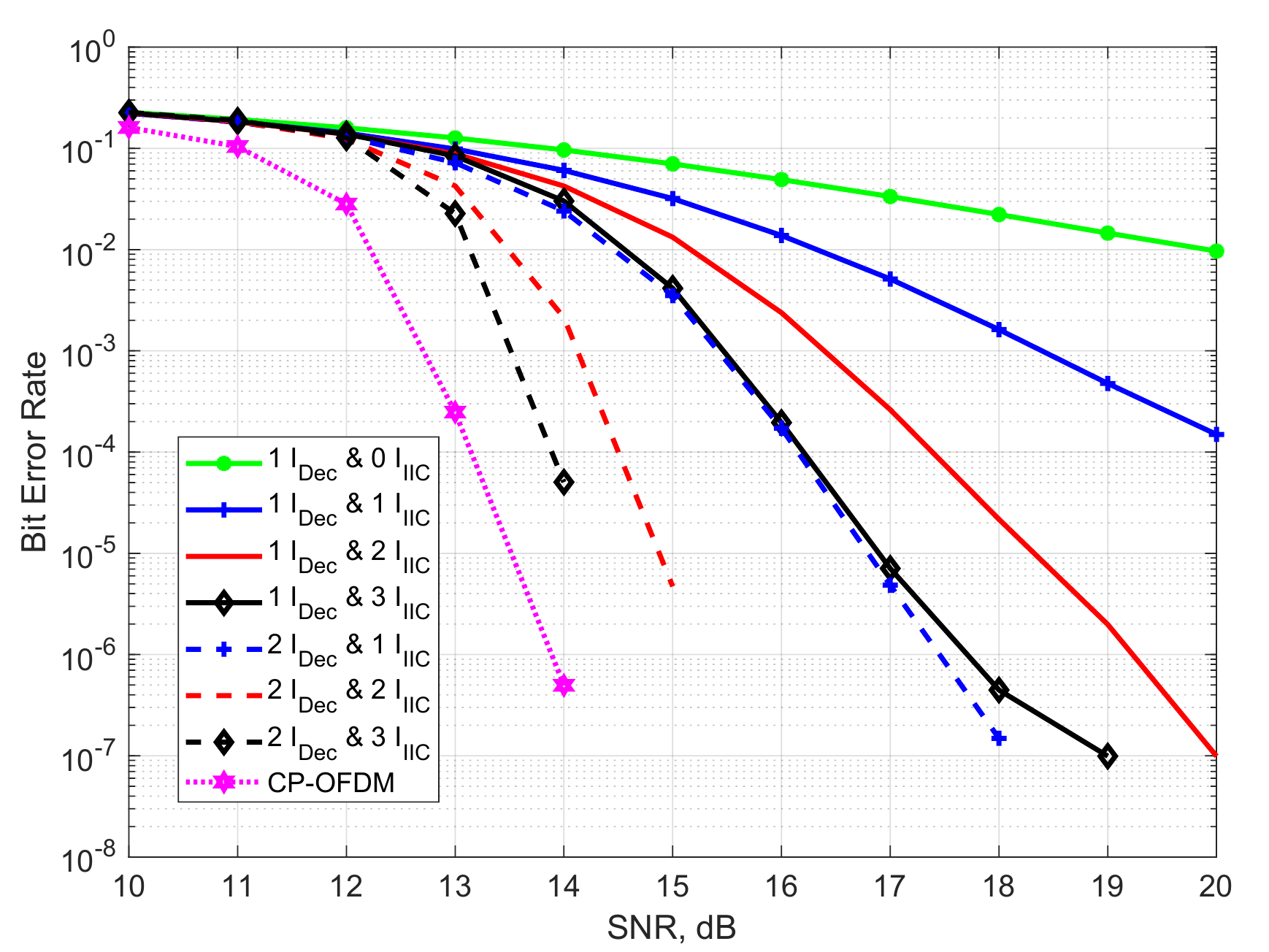}
    \caption{BER performance proposed receiver for FBMC-QAM and CP-OFDM}
  \end{subfigure}
\caption{Performance of the proposed IIC-based BICM-ID system with M\(16^r\) mapped 16-QAM over ETU channel} 
\end{figure}

Similarly, the performance of the proposed IIC-based BICM-ID receiver for FBMC-QAM and CP-OFDM over a single realization of the EVA and ETU channels are shown in Fig. 7 and Fig. 8, respectively. From Fig. 7(a), we can see that even though the EXIT curves at 14dB and \(\mathbb{I}_{Dec} = 1\) allow an open tunnel in the EPA channel (see Fig. 6(a)), they intersect in the EVA channel. This implies that either higher SNR values or more outer decoder iterations will be required to open a tunnel in the EVA channel. This is confirmed in Figs. 6(b) and Figs. 7(b) which show that with \(\mathbb{I}_{Dec} = 1 \) and \(\mathbb{I}_{IIC} = 2\), a BER of \(10^{-6}\) is obtained for the EPA channel at SNR of 16dB. However, an SNR of about 19.8dB is required to achieve a similar BER performance for the EVA channel. This is mainly due to the high frequency-selectivity of the EVA channel compared to EPA, which degrades the BER performance. Analogous to EPA, Fig. 7(b) and Fig. 8(b) show that once a tunnel has been created between the inner and outer decoder EXIT curves, increasing the number of IIC iteration can greatly improve the system performance by cancelling the interference caused by the non-orthogonality of FBMC-QAM and the high frequency selectivity of the EVA and ETU channels. Notice from  Fig. 7(b) and Fig. 8(b) that, for \(\mathbb{I}_{Dec} = 2 \) and \(\mathbb{I}_{IIC} = 3\) FBMC-QAM has 0.5dB and 1dB SNR loss compared to the benchmark CP-OFDM for EVA and ETU, respectively. 

\subsection{Performance of Multiple Fading Channel Realizations}
\label{SubSecResults2}
The above BER and EXIT curves have been shown for specific instances of the respective random fading channels. This is because the EXIT charts are defined for fading channels only if we assume very long codes and a channel that changes with the symbol rate. For multiple random instances of a frequency-selective channel, each realization will produce a different inner decoder EXIT curve. In order to generalize the use of EXIT chart in random fading channels, we propose a model that allows an outage probability of \( \rho \% \) in the EXIT chart analysis \cite{YahyaJHarbi:17}. That is, considering a \( \rho \% \) outage probability line on the EXIT curve, its convergence behaviour is expected to be satisfied by  \( (1-\rho) \% \) of the channel instances. As an example, the inner decoder EXIT curves for 50 random realizations of the EVA channel at SNR of 17dB is depicted in Fig. 9. The black lines represent the inner decoder EXIT curves for the individual realizations. Different levels of outage probability are also represented on Fig. 9. Note for instance the MI of 90th percentile line. It is expected that this MI can be achieved by \( 90 \% \) of the random fading channels (45 out of 50 in this case), with the remaining \( 10 \% \) in outage. Furthermore, if a tunnel exists between the 90th percentile line and the outer decoder EXIT curve, the decoding iterations will converge for \( 90 \% \) of channel instances.

\begin{figure}[t!]
\centering
\includegraphics[width=3.3in]{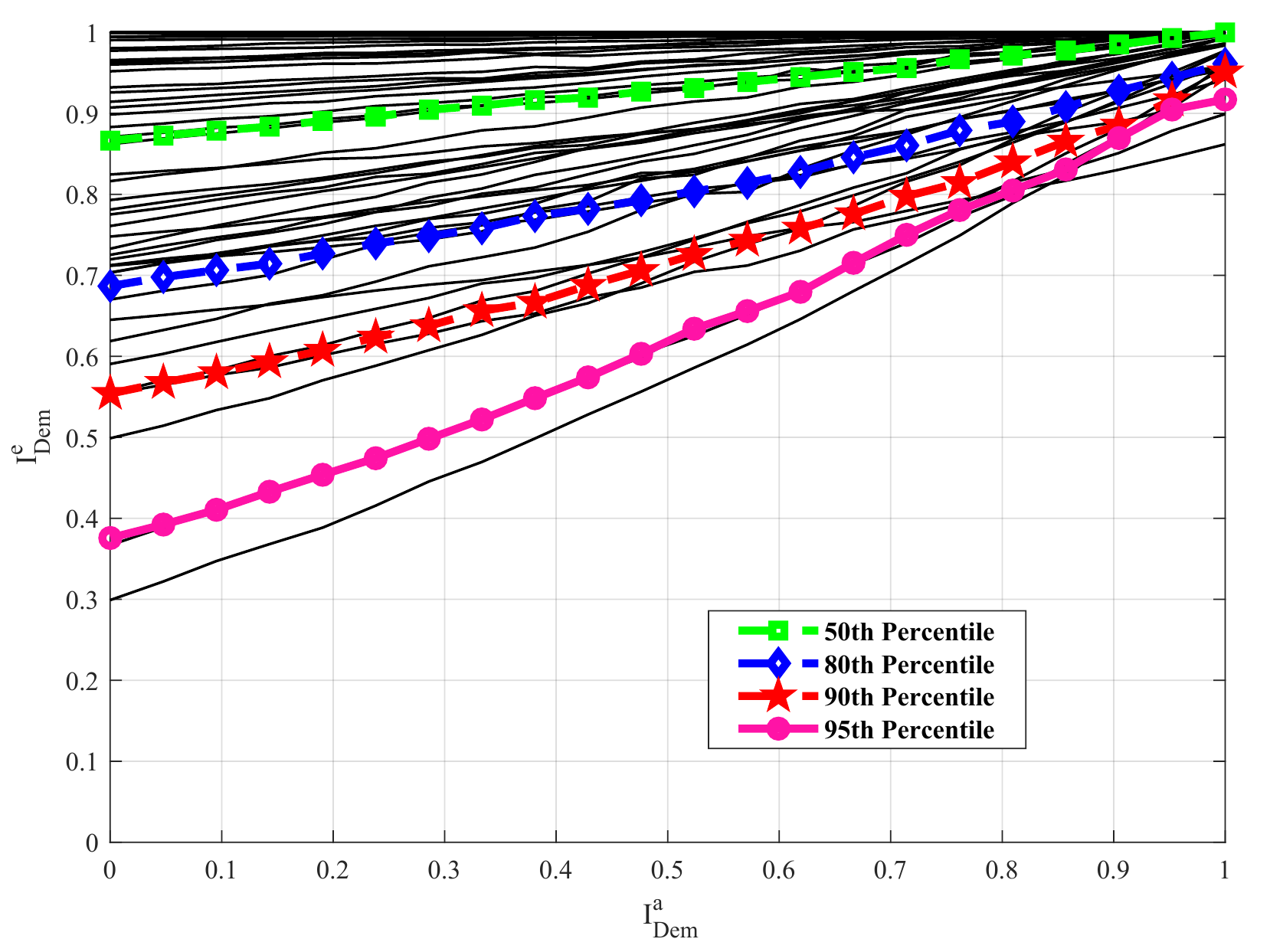}
\caption{Inner decoder EXIT curves for 50 random realizations of EVA channel showing different levels of outage probability.}
\end{figure}

\begin{figure}
  \begin{subfigure}{6cm}
    \centering\includegraphics[width=3.3in]{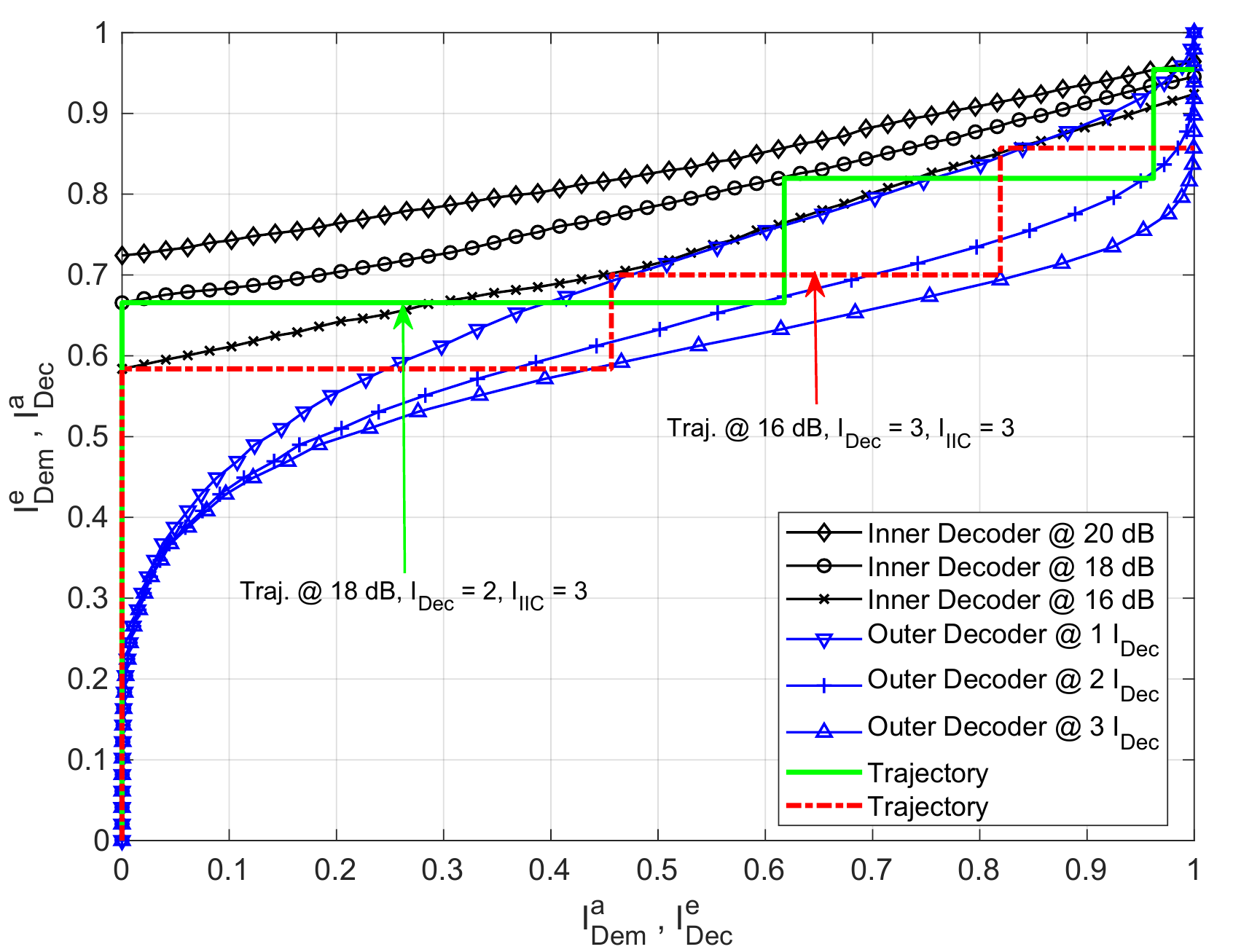}
    \caption{EXIT function and decoding trajectories}
  \end{subfigure}
  \begin{subfigure}{6cm}
    \centering\includegraphics[width=3.3in]{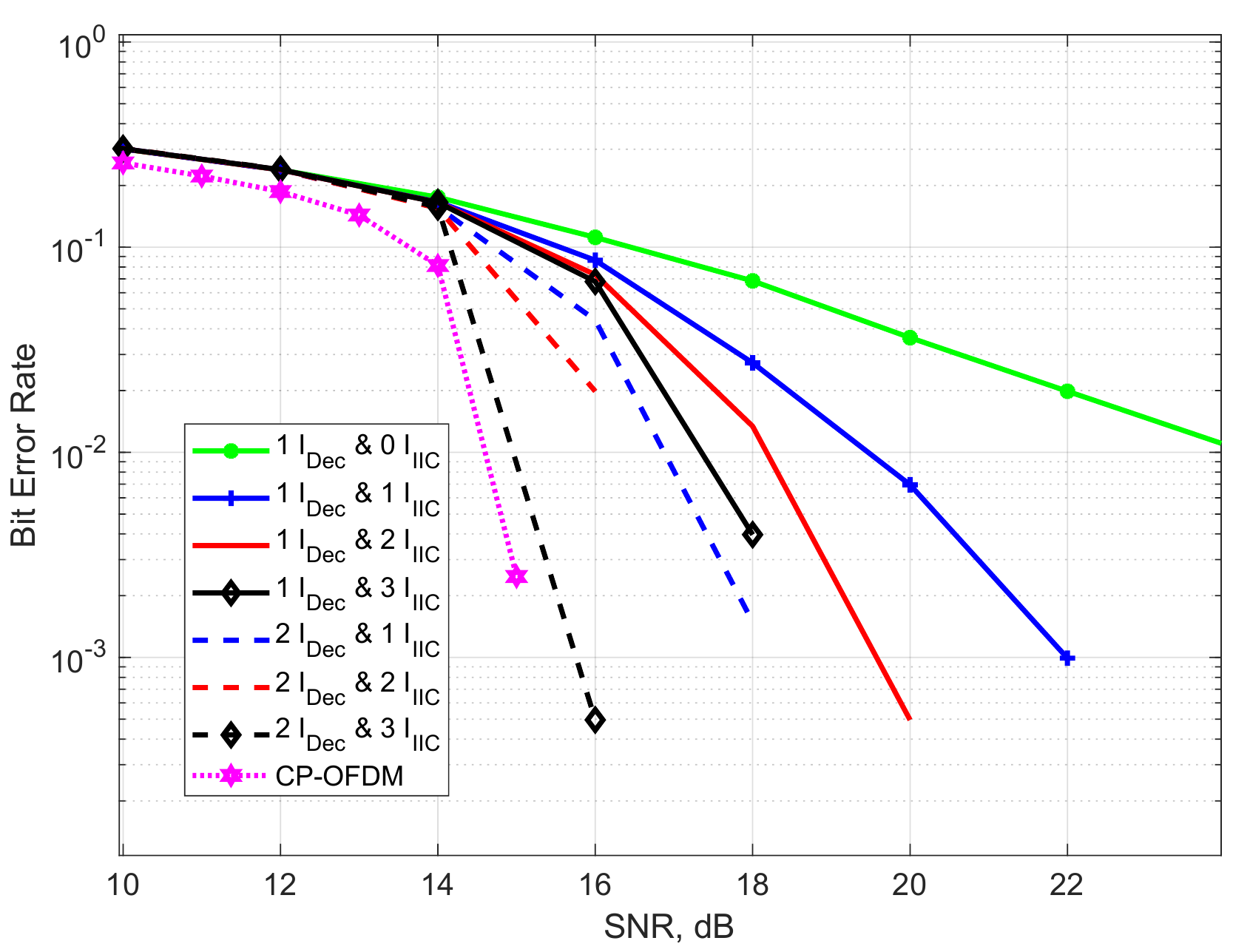}
    \caption{BER performance proposed receiver for FBMC-QAM and CP-OFDM}
  \end{subfigure}
\caption{Performance multiple realizations of EVA channel allowing a  \( 10 \% \) outage probability.} 
\end{figure}

Fig. 10(a) and Fig. 10(b) show the EXIT chart and BER performance, respectively, of the proposed BICM-ID receiver over 1000 realizations of the EVA channel, assuming \( 10 \% \) outage probability. This means that, 900 of the resulting inner decoder EXIT curves must satisfy the properties of the EXIT curve in Fig. 10(a). Notice that the inner decoder EXIT curve at 16dB intersects with the outer decoder EXIT curve using \(\mathbb{I}_{Dec} = 1 \). This implies that the tunnel between the two curves is blocked and therefore increasing the number of IIC iteration will not improve BER performance. By increasing \(\mathbb{I}_{Dec} \) to 2 and SNR to 18dB, a tunnel is created between the two EXIT curves as shown by the green trajectory line in Fig. 10(a). This translates to significant improvement in BER performance as shown in Fig. 10(b). Note that, in this case, the BER performance in Fig. 10(b) will be achieved for \( 90 \% \) of channel instances. Different levels of outage performance may be tolerated depending on the system requirements.

\subsection{Complexity Analysis}
This subsection presents the complexity evaluation of the proposed IIC-based BICM-ID receiver for FBMC-QAM compared with the benchmark CP-OFDM system. The overall complexity of the receiver depends on: (i) the modulator/demodulator processing in the inner decoder (ii) the VND and CND processing in the outer decoder and (iii) the number of outer decoder and IIC iterations. Mathematically, this complexity can be expressed as \cite{Chall:15}
\begin{equation}
\label{EQ28}
\begin{aligned}
\mathit{C}_{BICM-ID} &= \mathbb{I}_{Dec} \cdot \left(  \mathbb{I}_{IIC}+1 \right) \cdot \mathit{C}_{outer} \cdot N_b  \\ & + N_s \cdot [\mathit{C}_{inner,1} + \mathbb{I}_{IIC} \cdot \mathit{C}_{inner,i}]
\end{aligned}
\end{equation}
where \( \mathit{C}_{outer} \) denotes the complexity of the outer decoder, \( \mathit{C}_{inner,1} \) is the complexity of the first iteration of the inner decoder with no \textit{a priori} information from the outer decoder and \( \mathit{C}_{inner,i} \) is the complexity of the \(i\)-th iteration of the inner decoder considering the \textit{a priori} information from the outer decoder. \(N_b\) and \(N_s\) represent the number of information bits at the input of the LDPC encoder and number of symbols, respectively. Recall that \(\mathbb{I}_{Dec}\) and \(\mathbb{I}_{IIC}\) are the number of outer decoder and IIC iterations, respectively. \( \mathit{C}_{outer} \) depends on the scheduling used to exchange messages between the CND and VND. A detailed complexity derivation of the LDPC decoder is given in \cite{Chall:15} and has been omitted here for simplicity. Also, \( \mathit{C}_{inner,1} \) depends on the complexity of the multi-carrier (CP-OFDM or FBMC-QAM) demodulator during the initial decoder iteration whereas, \( \mathit{C}_{inner,i} \) is the combined complexity of the multi-carrier modulator and demodulator in subsequent decoder iterations. 

\begin{table*}[t!]
\label{Tab:3}
\caption{Complexity of IIC-based BICM-ID receiver for CP-OFDM and FBMC-QAM}
\begin{center} \vline
     \begin{tabular}{ p{2cm} | p{2cm} | p{2.2cm} | p{2cm} | p{2.5cm} | p{2.5cm}}
     \hline 
     \centering{ \textbf{IIC Iterations} } & \centering{\textbf{CP-OFDM} (\ref{EQ32})} & \centering{ \textbf{FBMC-QAM} (\ref{EQ33})} & \centering{ \textbf{Hybrid}} & \textbf{CP-OFDM/FBMC-QAM  Ratio} & \textbf{CP-OFDM/Hybrid Ratio}  \\ \hline
      0 & 448  & 960  & 960  & 2.14 & 2.14 \\ \hline
      1 & -    & 2880 & 2880 & 6.4  & 6.4  \\ \hline
      2 & -    & 4800 & -    & 10.7 & -      \\ \hline
      3 & -    & 6720 & -    & 15   & -      \\ \hline
     \end{tabular} 
\end{center} 
\end{table*} 

As mentioned in section \ref{Simulation:Setup}, there is negligible intrinsic interference in the benchmark CP-OFDM system due to the use of sufficient CP and guard band. Therefore, there is no IIC iterations for the benchmark CP-OFDM system. As a result, its complexity can be expressed as
\begin{equation}
\begin{aligned}
\label{EQ29}
\mathit{C}_{BICM-ID}^{OFDM} &= \mathbb{I}_{Dec} \cdot \mathit{C}_{outer} \cdot N_b + N_s \cdot \mathit{C}_{inner,1}^{OFDM} \\ & = \mathbb{I}_{Dec} \cdot \mathit{C}_{outer} \cdot N_b + N_s \cdot M/2 \log_2 (M)
\end{aligned}
\end{equation}
where \( \mathit{C}_{inner,1}^{OFDM} = M/2 \log_2 (M)\) is the complexity of the CP-OFDM demodulator, computed by considering the number of complex multiplications in the \(M\)-point FFT of OFDM \cite{Demmer:19}. On there other hand, due to the loss of orthogonality in FBMC-QAM there exist intrinsic interference. Therefore IIC is required to cancel the interference. Thus, the complexity of FBMC-QAM using the proposed iterative receiver is given as 
\begin{equation}
\label{EQ30}
\begin{aligned}
\mathit{C}_{BICM-ID}^{FBMC} & = \mathbb{I}_{Dec} \cdot \left(  \mathbb{I}_{IIC}+1 \right) \cdot \mathit{C}_{outer} \cdot N_b  \\ & + N_s \cdot \left(  1 + 2 \mathbb{I}_{IIC} \right) [M/2 \log_2 (M) + KM]
\end{aligned}
\end{equation}
\begin{proof}
The proof of (\ref{EQ30}) is given in Appendix. 
\end{proof}
Notice from Fig. 6(a) that there is no significant gain in MI of the outer decoder beyond 8 iterations. Therefore, as an example, we set the number of outer decoder iterations in the CP-OFDM system to \(\mathbb{I}_{Dec} = 8 \). For this example, (\ref{EQ28}) can be rewritten as 
\begin{equation}
\begin{aligned}
\label{EQ31}
\mathit{C}_{BICM-ID}^{OFDM} & = 8 \cdot \mathit{C}_{outer} \cdot N_b + N_s \cdot M/2 \log_2 (M)
\end{aligned}
\end{equation}
Moreover, as shown in Fig. 6(b), the FBMC-QAM system requires \( \mathbb{I}_{Dec} = 2 \) and \( \mathbb{I}_{IIC} = 3 \) to achieve a comparable BER performance with the benchmark CP-OFDM. In terms of complexity, this translates to
\begin{equation}
\label{EQ32}
\begin{aligned}
\mathit{C}_{BICM-ID}^{FBMC} &= 8 \cdot \mathit{C}_{outer} \cdot N_b + 7 N_s \cdot [M/2 \log_2 (M) + KM]
\end{aligned}
\end{equation}
From (\ref{EQ31}) and (\ref{EQ32}) it can be seen that the complexity of the outer decoder is similar for both CP-OFDM and FBMC-QAM. Therefore, the difference in complexity between CP-OFDM and FBMC-QAM is dominated by the structure of their respective modulator/demodulator and the number of IIC iterations. Hence, (\ref{EQ31}) and (\ref{EQ32}) can be simplified as 
\begin{equation}
\begin{aligned}
\label{EQ33}
\mathit{C}_{BICM-ID}^{OFDM} & = N_s \cdot M/2 \log_2 (M)
\end{aligned}
\end{equation}
and 
\begin{equation}
\label{EQ34}
\begin{aligned}
\mathit{C}_{BICM-ID}^{FBMC} &= 7 N_s \cdot [M/2 \log_2 (M) + KM],
\end{aligned}
\end{equation} respectively.

Using the simulation parameters in Table I ( \(K=4\) and  \( M = 128 \) ), the computational complexity of the two systems are shown in Table III. We can see that the number of complex multiplications increase with the number of IIC iterations, as expected. For \( \mathbb{I}_{IIC} = 3 \), which achieves similar BER performance as the CP-OFDM benchmark, the FBMC-QAM system requires 15 times more complexity than CP-OFDM. Note however that, the CP-OFDM benchmark consumes additional resources in terms of CP and guard band in order to maintain orthogonality and avoid intrinsic interference. Since mMTC applications are expected to connect billions of user terminals, the extra resources and signaling required to maintain synchronous communication between terminals may be prohibitive. Furthermore, OFDM has been shown to perform poorly in asynchronous communications \cite{YahyaJHarbi:16,CSexton:18}. This is as a result of the increased interference to adjacent subbands caused by the high spectrum leakage in OFDM. \cite{YahyaJHarbi:16} showed that in the case of insufficient CP and guard band residual interference exist in CP-OFDM and IIC iterations are needed to cancel this interference, which increases the complexity of CP-OFDM.

Also, as shown in Table III the complexity of the proposed receiver is dominated by the signal processing at the FBMC-QAM modulator and demodulator inside the inner decoder. Therefore, to reduce complexity of the BICM-ID receiver for FBMC-QAM, there exist the possibility that the interference reconstruction can be based on the output of the soft mapper. This would be equivalent to bypassing the FBMC-QAM modulator/demodulator in some IIC iterations. For example, consider a hybrid iterative system in which the first IIC iteration (\( \mathbb{I}_{IIC} = 1 \)) involves both the FBMC-QAM modulator and demodulator. However, in subsequent IIC iterations (\( \mathbb{I}_{IIC} > 1 \)), the interference cancellation is performed at the output of the soft mapper. This implies that, after the initial iteration without IIC and the first IIC iteration, the complexity of the hybrid iterative receiver is independent of the signal processing in the FBMC-QAM modulator/demodulator. From the example shown in Table III, it can be seen that the ratio of complexity of the iterative receiver to that of CP-OFDM saturates at about 6.4 and increasing the number of IIC iterations will not significantly increase the complexity. Thus, the hybrid receiver can serve as a suboptimal approach that is capable of providing a trade-off between complexity and BER performance of the proposed iterative receiver.

%The complexity of the hybrid iterative system can therefore be expressed as
%\begin{equation}
%\label{EQ35}
%\begin{aligned}
%\mathit{C}_{BICM-ID}^{Hybrid} &=  N_s \cdot \{ 3 [ M/2 \log_2 (M) + KM ] + \mathbb{I}_{IIC} - 1 \}
%\end{aligned}
%\end{equation}

\section{Conclusion}
In this paper, FBMC-QAM has been studied, as an alternative to conventional CP-OFDM, due to its high spectral efficiency and ultra-low OOB emission, which make it suitable for future applications in 5G such as asynchronous mMTC. The key issue with FBMC-QAM is the high intrinsic interference caused by the loss of complex orthogonality between subcarriers. To address the interference problem we propose and analyse the performance of an IIC-based BICM-ID receiver for FBMC-QAM systems. We studied the convergence behaviour of the iterative receiver using EXIT charts. Based on the EXIT chart analysis, the number of iterations of the proposed BICM-ID receiver required to achieve a target BER performance and the corresponding complexity is predicted. The results show that the proposed IIC-based LDPC BICM-ID receiver can remove the intrinsic interference in FBMC-QAM systems under time-varying channels with an increase in computational complexity compared to CP-OFDM. For example, in the case of the EPA channel, FBMC-QAM with IIC requires about 15 times more complexity (due to the inner modulator/demodulator) than in a benchmark CP-OFDM system with no IIC. However, the EXIT chart analysis can be used to minimise complexity for given performance by choosing the optimum number of inner and outer iterations. Also, there is the potential for a hybrid receiver that bypasses the FBMC-QAM modulator/demodulator after the initial IIC iteration to be used as a low complexity implementation of the proposed receiver. In summary, FBMC-QAM is a promising multicarrier waveform for asynchronous mMTC applications due to its ultra-low OOB emission, which causes very low leakage interference between asynchronous users, and the proposed iterative receiver is capable of effectively addressing the intrinsic interference problem. 
%--------------------------------------------------------------------------------------
% APPENDIX
%--------------------------------------------------------------------------------------
%\appendices
% APPENDIX A
%--------------------------------------------------------------------------------------
\section*{Appendix \\ Proof of Equation (\ref{EQ30})}
\label{App:App1}
Using (\ref{EQ28}), the complexity of the iterative receiver for FBMC-QAM is given as
\begin{equation}
\label{EQ35}
\begin{aligned}
\mathit{C}_{BICM-ID}^{FBMC} &= \mathbb{I}_{Dec} \cdot \left(  \mathbb{I}_{IIC}+1 \right) \cdot \mathit{C}_{outer} \cdot N_b  \\ & + N_s \cdot [\mathit{C}_{inner,1}^{FBMC} + \mathbb{I}_{IIC} \cdot \mathit{C}_{inner,i}^{FBMC}]
\end{aligned}
\end{equation}
where \( \mathit{C}_{inner,1}^{FBMC} \) is the complexity of the FBMC-QAM demodulator in the first iteration and \( \mathit{C}_{inner,i}^{FBMC} \) is the combined complexity of the FDMC-QAM modulator and demodulator in the subsequent iterations. In terms of complex multiplications FBMC-QAM requires \(KM\) multiplications for the filtering process and followed by an \(M\)-point FFT. Therefore, 
\begin{equation}
\label{EQ36}
\mathit{C}_{inner,1}^{FBMC} = M/2 \log_2 (M) + KM 
\end{equation}
and
\begin{equation}
\label{EQ37}
\mathit{C}_{inner,i}^{FBMC} = 2(M/2 \log_2 (M) + KM).
\end{equation}
Note that the factor 2 takes into consideration the fact that both FBMC-QAM modulation and demodulation processes are performed during the \(i\)-th IIC iteration. Substituting  (\ref{EQ36}) and (\ref{EQ37}) into the second term of (\ref{EQ35}) gives
\begin{equation}
\label{EQ38}
\begin{aligned}
\mathit{C}_{inner}^{FBMC} &= \mathit{C}_{inner,1}^{FBMC} + \mathbb{I}_{IIC} \cdot \mathit{C}_{inner,i}^{FBMC} \\ &= M/2 \log_2 (M) + KM \\ & + \mathbb{I}_{IIC} \cdot [2(M/2 \log_2 (M) + KM)] \\ & = M/2 \log_2 (M) + \mathbb{I}_{IIC} \cdot M \log_2 (M) \\ & + KM + 2 \mathbb{I}_{IIC} KM \\ & = M/2 \log_2 (M) (1 + 2 \mathbb{I}_{IIC}) + KM (1 + 2 \mathbb{I}_{IIC})
\end{aligned}
\end{equation}
Rearranging the last term of (\ref{EQ38}), the complexity of the inner decoder due to the FBMC-QAM modulator/demodulator can be finally expressed as 
\begin{equation}
\label{EQ39}
\mathit{C}_{inner}^{FBMC} = (1 + 2 \mathbb{I}_{IIC}) [M/2 \log_2 (M) + KM]
\end{equation}
(\ref{EQ30}) follows straightforwardly from substituting (\ref{EQ39}) into (\ref{EQ35}). $\square$ 
% \\ & = \mathbb{I}_{Dec} \cdot \left(  \mathbb{I}_{IIC}+1 \right) \cdot \mathit{C}_{outer} \cdot N_b  \\ & + N_s \cdot \left(  1 + 2 \mathbb{I}_{IIC} \right) [M/2 \log_2 (M) + KM]

\bibliographystyle{IEEEtr}
\bibliography{Spotlight}

\end{document}